\documentclass[%
 aip,
 pra,
 amsmath,amssymb,
 preprint,%
]{revtex4-2}

\usepackage{tikz}
\usepackage{amsmath}
\usetikzlibrary{decorations.pathreplacing}
\usepackage{amssymb,amsmath,amsfonts}
\usepackage{graphicx,graphics}
\usepackage{epsfig}
\usepackage{blkarray}
\usepackage{epstopdf}
\usepackage[FIGBOTCAP]{subfigure}
\usepackage{amsfonts}
\usepackage{bm,bbm}
\usepackage{amssymb,amsmath,amsfonts}
\usepackage{mathrsfs}
\usepackage{calc}
\usepackage{epsfig}
\usepackage{float}
\usepackage{color}
\usepackage{epstopdf}
\usepackage{bm,amssymb}
\usepackage{graphicx,color}
\usepackage{comment}
\usepackage{soul}
\usepackage{amsmath}
\newcommand{\be}{\begin{equation}}
\newcommand{\ee}{\end{equation}}

\newcommand{\beqar}{\begin{eqnarray}}
\newcommand{\eeqar}{\end{eqnarray}}
\newcommand{\bcen}{\begin{center}}
\newcommand{\ecen}{\end{center}}

\DeclareUnicodeCharacter{2212}{-}

\begin{document}
\title{Accelerated Hydrogen Exchange Reaction in a Dark Cavity: A Benchmark for Bridging the Gap Between Theory and Experiment }

\author{Victor Berenstein$^1$, Giacomo Valtolina$^4$, Zohar Amitay$^1$, and  Nimrod Moiseyev$^{1,2,3}$ \footnote{\tt nimrod@technion.ac.il https://nhqm.net.technion.ac.il} }

\affiliation{$^1$ Schulich Faculty of Chemistry, Faculty of Physics$^2$ and Solid State Institute$^3$, Technion-Israel Institute of Technology, Haifa 32000, Israel
\newline
$^4$Fritz-Haber-Institut der Max-Planck-Gesellschaft, Faradayweg 4-6, 14195 Berlin, Germany
}

\begin{abstract}
{The gas-phase hydrogen exchange reaction (HER) is the most fundamental chemical
process for benchmarking quantum reaction dynamics. In this Letter, we focus on
controlling HER by means of strong light-matter coupling inside a resonant cavity,
an approach often called polariton chemistry. In particular, we focus on the isotopic
variation of HER involving collisions between molecular hydrogen H$_2$ and deuterium
atom D, i.e., H$_2$+D$\to$HD+H. We find that the asymmetry introduced by the
different isotopes, despite being small, enables strong cavity-induced modifications of
reaction rates. Outside of the cavity the reaction is as usual D+H${_2}$$\to$DH+H.
However, inside the cavity another type of reactions take place where D+H$_2$$\to$DH+H+E$_{photon}$, where E$_{photon}$=$\hbar\omega_{cav}$. 
%
% By measuring the difference in kinetic energy and electronic and vibrational 
% energies between the reactants and the products and showing that it matches the
% cavity photon energy, one can prove the existence of the new type of reaction
% induced by the cavity, 
% and compare the corresponding theoretical and experimental reaction rates.
%
%It also allows us to compare the computed reaction rates in the cavity with the experimental results depending on the density of molecules in the cavity. 
%
Our results show that HER is an ideal platform to 
%finally 
make a significant step toward closing the gap between theory and experiment in polariton chemistry.}
\end{abstract}

\maketitle

\section{Introduction}

There have been many formal and informal discussions about the effects of a dark cavity on chemical reaction rates in the single-molecule level 
and, if exist, whether they remain also when 
many molecules reside in the cavity.
%the molecular density in the cavity is high.
See for example Ref.\citenum{lindoy2024investigating}.

The gas-phase hydrogen exchange reaction (HER) is the most fundamental chemical reaction. %process for benchmarking quantum reaction dynamics. 
HER benefits from exceptionally accurate theoretical modeling\cite{siegbahn1978accurate}. The small number of degrees of freedom and well-characterized potential energy surfaces make it a benchmark system for quantum reaction dynamics\cite{Xie2020hd}. In addition, the isotopic variation of HER,
 the reaction $H_2 + D \rightarrow HD + H$ 
 involving collisions between molecular hydrogen $H_2$ and deuterium  {atom} $D$, fulfills the two main conditions for cavity-modification of chemistry highlighted in Ref.\citenum{moiseyev2024conditions}, that is (1) the reaction must be asymmetric and associated with a single saddle point along the reaction coordinate $X$; and (2) the frequencies of the normal modes perpendicular to the reaction coordinate must vary with $X$. 
 
 In this Letter, we focus on controlling HER by means of strong light-matter coupling inside a resonant cavity, an approach often called polariton chemistry. 
 In particular, using recent theoretical progress \cite{moiseyev2024conditions},
 we focus on the isotopic variation of HER, 
 %involving collisions between molecular hydrogen $H_2$ and deuterium  {atom} $D$, i.e., 
 $H_2 + D\rightarrow HD+H$. We find that the asymmetry introduced by the different isotopes, despite being small, enables strong cavity-induced modifications of reaction rates. { Outside of the cavity the reaction is as usual $D+H_2\to DH+H$. However, inside the cavity another type of reactions take place where $D+H_2\to DH+H +\hbar\omega_{photon}$. 
 %
%  By measuring the difference in kinetic energy and electronic and vibrational 
% energies between the reactants and the products and showing that it matches the
% cavity photon energy, 
%
By measuring the emitted photon(s) in correlation to the measurements of the reaction products,
one can prove experimentally the existence of the new type of reaction induced by the cavity, 
and compare the corresponding theoretical and experimental reaction rates. 
This would be a significant step toward closing the gap between theory and experiment in polariton chemistry.

The present study is conducted within   
the non-Hermitian quantum framework~\cite{NHQM-BOOK},
and our theoretical model accounts for  
the HER reaction taking place between 
each pair of reactants, $D$ and $H_{2}$,  
by accounting for
two distinct resonance % transition 
states of the $DHH$ molecule through which the reaction can take place. 
The eigenvalue of the non-Hermitian complex-scaled Hamiltonian 
corresponding to each resonance state is complex, 
where the real part is the energy of the state,
and the imaginary part is proportional to the rate of the reaction when it takes place via this state.
The full system of HER reaction(s) taking place inside a cavity 
is then analyzed 
using these two resonance states, within 
the Jaynes-Cummings model~\cite{JaynesCummingsModel} for the case of a single DHH molecule in the cavity (i.e., a cavity with a single reactants-products system),
and within the 
Tavis-Cummings~\cite{TavisCummingsModel} model for the case of several and many DHH molecules in the cavity (i.e., a cavity with multiple reactants-products systems).
Initially, only one of the resonances—the one that is energetically higher—is populated in each DHH molecule.
Hence, without a cavity, the HER reaction proceeds exclusively via this state.
The second resonance state becomes populated,
along with the emission of a cavity photon,
only due to the interaction between the molecule and the cavity field and the resulting formation of the 
system's polaritonic states.
This is effectively achieved when the cavity frequency matches the energy difference between the real parts of the two resonance states.
Then, the HER reaction also takes place via this second resonance state.
As a result, for a given pair of such molecular resonance states, each corresponding to a different individual reaction rate, 
the overall reaction rate in a cavity may be either enhanced or suppressed 
relative to the rate without a cavity, 
generally depending on 
the coupling strength between the cavity and 
an individual molecule and
on the number of molecules in the cavity. 
This mechanism of cavity-induced reaction rate 
modification is active even when the cavity is initially dark, i.e., when it initially contains no photons.

Before delving into the details of the study, 
we first present in Fig.~\ref{ENHANCEMENT} 
the final single-molecule results of our calculations, obtained with a properly selected pair of transition states of the DHH molecule and 
an appropriately chosen cavity. 
It shows the results for the enhancement of the reaction rate due to the initially-dark cavity for the HER reaction D+H$_{2}$$\rightarrow$DH+H,
with a single excited transition-state molecule in the 
cavity.
Shown is the cavity-induced rate enhancement,
$\text{R}_{\text{cav,no-cav}} = \Gamma_{\text{cav}} / \Gamma_{\text{no-cav}}$,
as a function of the 
coupling parameter (strength) between the cavity and 
a single molecule, 
$\epsilon_{\text{cav}}$.
The quantities $\Gamma_{\text{cav}}$ and $\Gamma_{\text{no-cav}}$ denote 
the reaction rates with and without the cavity (i.e, inside and outside the cavity), respectively.
As seen, a considerable cavity-induced enhancement of the reaction rate is observed here in the 
initially-dark cavity for the single-molecule case.
Later in the paper,
we show that %, 
%for a cavity-molecule coupling parameter 
%in the range of 0.005 to 0.020 a.u.,
such considerable enhancement 
is maintained also in the 
several- and many-molecule cases.
%
%We have not been able to obtain results for smaller values of the coupling parameter due to numerical computational limitations.
%
%Later, we will show that the enhancement of the reaction rate by the cavity does not necessarily disappear when we solve the Tavis–Cummings model for N molecules in the cavity \cite{TavisCummingsModel} rather than the Jaynes-Cummings model for a single molecule\cite{jaynes2005comparison}}.
%
The specific pair of resonance states 
considered in the calculations of Fig.~\ref{ENHANCEMENT} are as follows.
One resonance state, which is of the higher 
energy, is a conventional reaction transition state, 
where the HER reaction takes place 
at an energy close to the height of the reactants-to-products potential barrier.
This is the initially-populated state.
In contrast, the other resonance state corresponds 
to a non-conventional reaction pathway, 
in which the HER reaction takes place 
through a dynamical barrier that is associated with the energy transfer (mediated by non-adiabatic couplings)
from the reaction coordinate to the bound normal modes perpendicular to the reaction coordinate. 
In a collinear reaction A+BC$\to$AB+C, the reaction coordinate is the  asymmetric normal mode, where the normal bound modes 
%that are 
perpendicular to the reaction coordinate are the symmetric and the two bending normal bound 
vibrations~\cite{moiseyev2024conditions}.

%----------------------------------------

\begin{figure} %[t!]
\centering
%\hspace*{+0.9cm}
\includegraphics[angle=000,scale=0.45]
{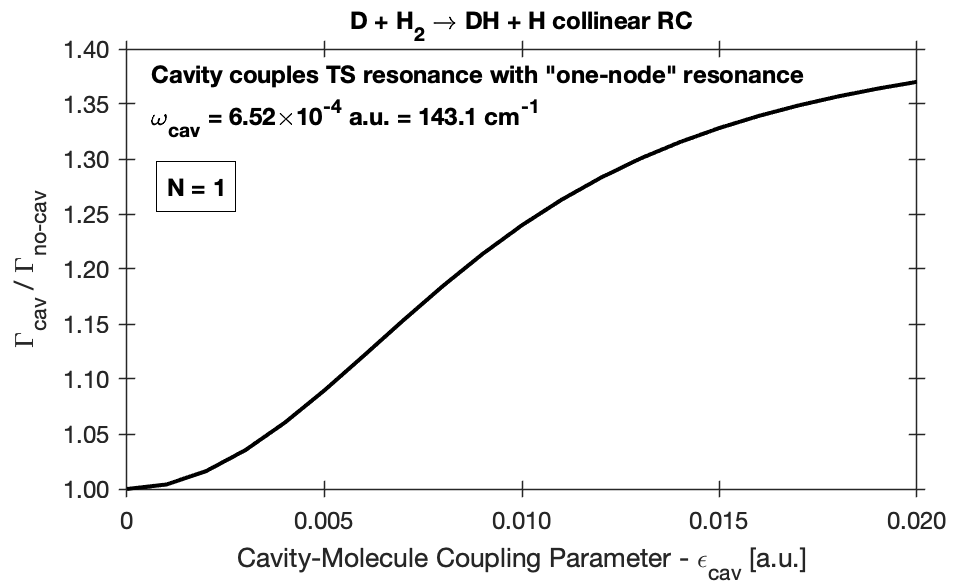}
\caption{
Calculated results for the enhancement of the reaction rate due to the initially-dark cavity for the HER reaction D+H$_{2}$$\rightarrow$DH+H,
with a single excited transition-state molecule in the 
cavity.
Shown is the cavity-induced rate enhancement,
$\text{R}_{\text{cav,no-cav}} = \Gamma_{\text{cav}} / \Gamma_{\text{no-cav}}$,
as a function of the 
coupling parameter (strength) between the cavity and 
a single molecule, 
$\epsilon_{\text{cav}}$ (given in atomic units).
The quantities $\Gamma_{\text{cav}}$ and $\Gamma_{\text{no-cav}}$ denote 
the reaction rates with and without the cavity (i.e, inside and outside the cavity), respectively.
}
\label{ENHANCEMENT}
\end{figure}

%--------------------------------------------

The outline of the paper is as follows. First, we describe how the potential energy surface for the $D + H_2 \rightarrow DH + H$ reaction is calculated using mass-weighted nuclear coordinates. 
Then, we explain how the vibrational frequencies of the bending and symmetric stretching normal modes are computed as functions of the asymmetric vibrational normal mode (the reaction coordinate). %, $X_{RC}$. 
We also discuss the calculation of the reaction's potential barrier and the adiabatic one-dimensional (1D) potential that incorporates the ground-state energy of the bound vibrational modes. 
Next, we compute the complex pole resonances of the adiabatic 1D potential to determine the reaction rate outside the cavity, and explain how this rate is calculated.
Subsequently, 
we describe how the reaction rate inside a given initially-dark cavity is calculated in the case of a single molecule in the cavity, and present the corresponding results shown of Fig.~\ref{ENHANCEMENT}.
We then extend the theoretical framework and present results for the case of many molecules in the cavity.
Lastly, we outline how experiments could be conducted to study polariton chemistry 
in gas-phase bimolecular collisions,  
with the goal of providing experimental evidence to the cavity-induced enhancement mechanism presented in this work.

%-----------------------------

%\newpage

\section{Calculations of the potential energy surface for $D+H_{2}\to DH+H $ reaction using the mass weighted nuclear coordinates}

In Fig.~\ref{COUNTERPLOT} the contour plot shows the potential energy surface of $DHH$ for the collinear reaction as a function of two mass-weighted  coordinates $R_1^{MW}$ and $R_2^{MW}$ is presented. $R_1$ is the distance between the deuterium atom and the center of mass of the $H_2$ molecule. $R_2$ is the distance between the two hydrogen atoms in $H_2$. See below the schematic representation of $R_1$ and $R_2$ in the collinear collision of $D$ with $H_2$. The mass-weighted coordinate $R_1^{MW}$ is defined as $R_1\sqrt{\mu_{DHH}/m_D}$, where the reduced mass $\mu_{DHH}$ is given by $\mu_{DHH} = [1/m_D + 1/{2m_{H}}]^{-1}$, and $m_D$ and $m_H$ are the masses of deuterium and hydrogen, respectively. The mass-weighted coordinate $R_2^{MW}$ is defined as $R_2\sqrt{\mu_{HH}/m_D} = R_2\sqrt{m_H}/2$. The saddle point is associated with the transition state (TS). The potential energy barrier along the reaction coordinate $X_{RC}$ has a maximum value at TS. The reaction coordinate is associated with the asymmetric stretching normal mode. The normal bound stretching mode and the two normal degenerate bound bending modes are perpendicular to the reaction coordinate. 
{
%$\mu_{D-H_2}=\left(\frac{1}{m_D}+\frac{1}%{m_{H_2}}\right)^{-1}$
%$\mu_{H_2}=\left(\frac{1}{m_H}+\frac{1}{m_{H}}\right)^{-1}$

\begin{center}
\begin{tikzpicture}[thick]

  % Nodes
  \filldraw (0,0) circle (2pt) node[below=2pt] {\textbf{D}};
  % Circle for H(1)
  \filldraw (6.0,0) circle (2pt);
  % Label for H(1) shifted right
  \node at (6.2, -0.3) {\textbf{H}$^{\boldsymbol{(1)}}$};

  % Circle for H(2)
  \filldraw (8.0,0) circle (2pt);
  % Label for H(2) shifted left
  \node at (8.2, -0.3) {\textbf{H}$^{\boldsymbol{(2)}}$};

  % Line
  \draw[black, line width=1pt, dashed] (-1,0) -- (9,0);

  % Distances
  \node at (3.5, -0.3) {\boldsymbol{$r_1$}};
   % Label for R(2)=r(2) shifted left and up
  \node at (7, +0.25) {$\boldsymbol{R_2 = r_2}$};

  % Brace for R1
  \draw[decorate,decoration={brace,amplitude=10pt},yshift=11pt]
     % Label for R(1) shifted up by 17 pt
    (0,0) -- (7,0) node[midway,yshift=17pt] {\boldsymbol{$R_1$}};

\end{tikzpicture}

\begin{figure}[h!]
     \centering
     \hspace*{+0.9cm}
      \includegraphics[angle=000,scale=0.5]{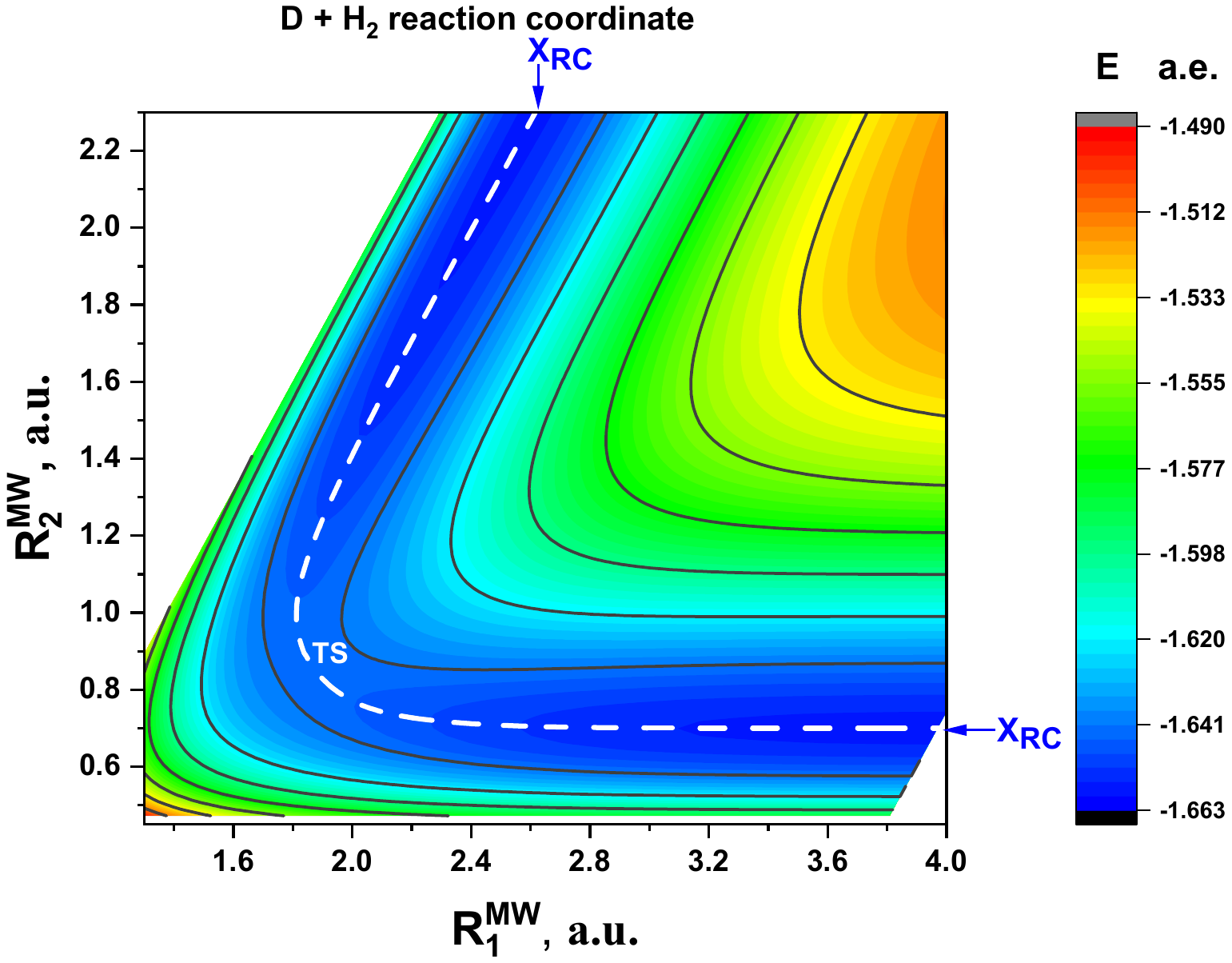}
     \caption{Contour plot of the potential energy surface of $DHH$ for the collinear reaction as a function of two mass weighted   coordinates  $R_1^{MW}$ and $R_2^{MW}$. }
    \label{COUNTERPLOT}
    \end{figure}
\end{center}
%    {\color{red} VICTOR REFERENCES SHOULD BE MENTIONED AS USUAL IN PAPERS}
%    \\
    {
    We selected the most energetically favorable~\cite{siegbahn1978accurate,klupfel2012effect,jankunas2014simplest} reaction pathway with the collinear approach of D toward $H_2$.  
DFT calculations were carried out at the mPW2PLYP/6-311++G(d,p) level employing the Gaussian 16 quantum chemistry package~\cite{g16}. The mPW2-PLYP functional~\cite{schwabe2006towards,schwabe2007double}  is derived from mPW exchange functional in conjunction with LYP correlation. This functional has been shown to yield improved energetics and a narrower error distribution than B3LYP. Potential energy surface, see Fig.2, was calculated scanning both interatomic distances at the range between 3.0 \AA~(5.7 a.u.) to 0.5 \AA~ (0.9 a.u.) with an increment of 0.02 \AA~(0.04 a.u.). Also, intrinsic reaction coordinate (IRC), shown by the dashed line on Fig.2, mass-weighted force constant matrix, and the corresponding normal modes (NM) vibrational frequencies were calculated at the same range and step size. The computed geometries and energies are consistent with previous literature results~\cite{klupfel2012effect,su2015h+,jankunas2014simplest}.
    }
    
\section{Frequencies of Bending and Symmetric Stretching Normal Modes Along the Reaction Coordinate for $D+H2\to DH+H$ reaction} 

{ In Fig.~\ref{1DadiabaticPEC}, the one-dimensional (1D) potential energy barriers for the reaction $D + H_2 \to DH + H$, as obtained from our calculations, are presented. The potential barrier, $V_{rc}(X_{RC})$, is derived by associating the reaction coordinate with the linear curved coordinate indicated by the dashed white line in Fig.~2.

Let us explain how the calculations were performed. For sufficiently large values of $R_1^{MW}$, the coordinate $R_2^{MW}$ adopts a value corresponding to the equilibrium bond length of the $H_2$ molecule multiplied by $\sqrt{1/2}$, where the potential energy reaches a minimum as $R_2^{MW}$ is varied. As $R_1^{MW}$ decreases, we track the value of $R_2^{MW}$ that minimizes the potential energy, continuing this process until reaching the product region, where both $R_1^{MW}$ and $R_2^{MW}$ become large.

The potential energy along this reaction coordinate yields the potential energy barrier, as illustrated in Fig.~3. The adiabatic potential, defined as $V_{ad}(X_{RC}) = V_{rc}(X_{RC}) + \hbar\Omega(X_{RC})/2$,
 is obtained under the assumption that the motion along the reaction coordinate is much slower than the motions associated with the bound vibrational bending and symmetric stretching normal modes, which are perpendicular to the reaction coordinate.

The frequencies of these vibrational modes are derived from the eigenvalues of the mass-weighted Hessian matrix, which describes the second derivatives of the potential energy with respect to nuclear displacements. Mathematically, for each eigenvalue $\lambda$ of the Hessian matrix, the corresponding vibrational frequency $\nu$ is given by $\nu = \sqrt{\frac{\lambda}{m}}$,
where $m$ is the reduced mass associated with the respective normal mode. Positive eigenvalues correspond to real frequencies, indicating stable vibrations near a local minimum on the potential energy surface (PES). In contrast, negative eigenvalues lead to imaginary frequencies, signaling a direction of negative curvature associated with a saddle point on the PES. These imaginary-frequency modes align with the reaction path and are typically excluded from the adiabatic zero-point energy (ZPE) correction, which only includes vibrational modes perpendicular to the reaction coordinate.}

\begin{figure}[h!]
     \centering
     \hspace*{+0.9cm}
      \includegraphics[angle=000,scale=0.5]{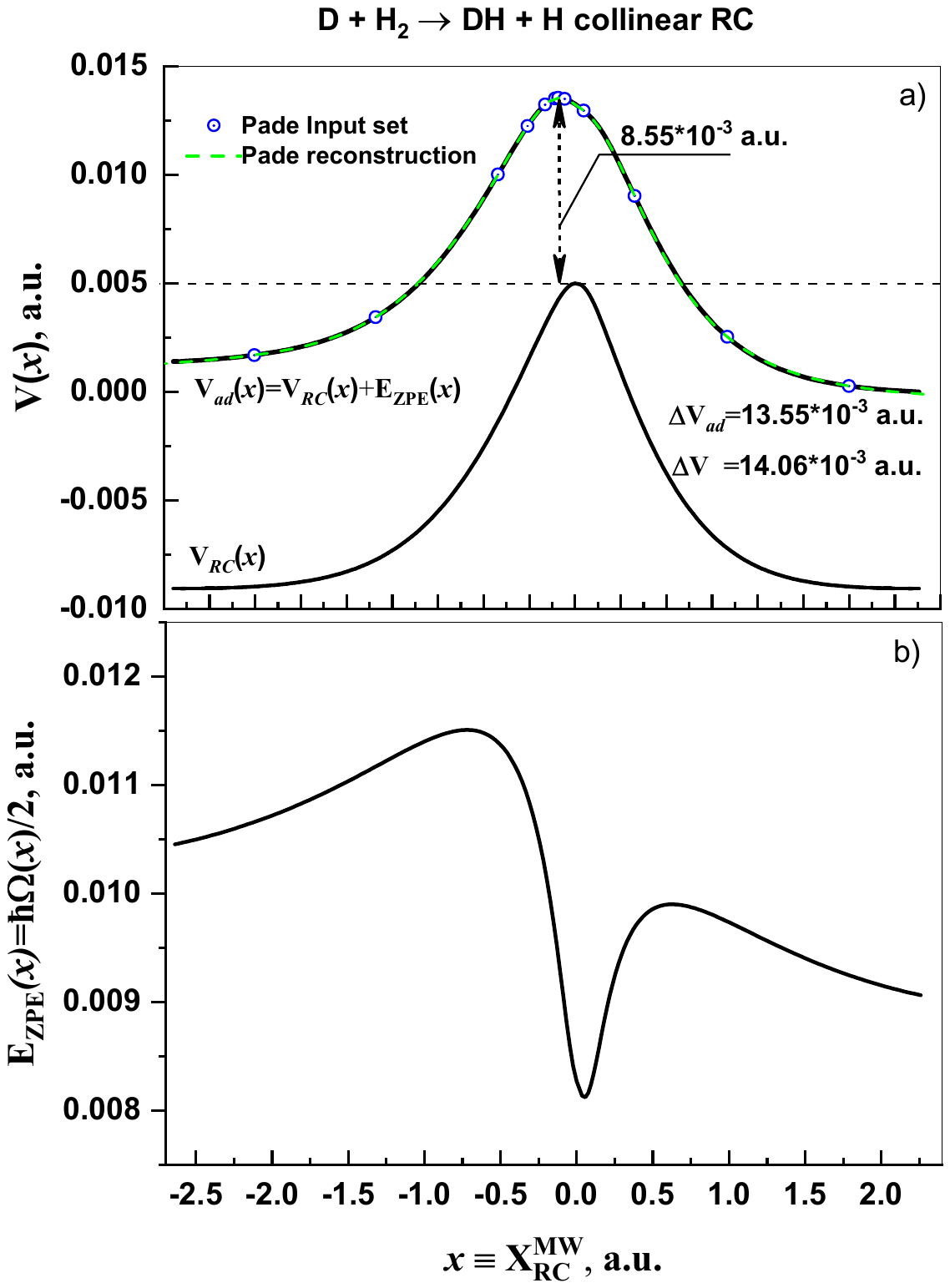}
     \caption{    The one-dimensional (1D) potential barrier for the reaction $D+H_2\to DH+H$
is shown. The solid black line represents the potential barrier, $V_{rc}(X_{RC})$, as a function of the reaction coordinate ({associated with} the asymmetric stretching mode), 
The dashed green line shows the { Padé reconstruction of the }adiabatic potential, $V_{ad}(X_{RC})=V_{rc}(X_{RC})+\hbar\Omega(X_{RC})/2$, where $\Omega(X_{RC})$ 
 is the sum of the frequencies of the degenerate bending bound normal modes and the bound symmetric stretching normal mode.}
    \label{1DadiabaticPEC}
    \end{figure}

The numerical results marked by open circles in Fig.\ref{1DadiabaticPEC} were used as data for the presentation of the adiabatic potential $V_{ad}(X_{RC})$ as a ratio between two polynomials using the Schlessinger approach to Padé approximation.

In the Padé approach, the $V_{ad}(X_{RC})$ function is represented as a ratio of two polynomials.
\begin{equation}
V_{ad}(X_{RC}) = \frac{P(X_{RC})}{Q(X_{RC})}
\label{PADE_ratio}
\end{equation}
{Note that the equality of the ratio between two polynomials holds only when ${ V_{ad}(X_{RC}) }$ is an analytic function.}
 For example, if there are singular points at $\{X_{RC}=X_{J_{singular}}\}_{J_{singular}=1,2,..,M}$, then Eq.~\ref{PADE_ratio} holds only within the intervals $X_{J_{singular}-1}>X_{RC}>X_{J_{singular}}$~\cite{masjuan2013pade}.

However, the Schlesinger method~\cite{schlessinger1968use} overcomes this problem and effectively describes functions wherever they are analytic and exhibits non-regular behavior at singular points.  Namely, Schlesinger method works well every where beside at the singular points. This capability has already been used to calculate energy thresholds and resonance decay poles from experimental data on $\pi\pi$ scattering and the cross-section ratio in hadronic collisions~\cite{TRIPOLT2017411}.

The Padé like expansion of  Schlessinger ~\cite{schlessinger1968use} is given below,

\begin{eqnarray}
\label{SCHLESSINGER}
V_{ad}(X_{RC}) = \cfrac{V_{ad}(X_{RC_1})}{1+\cfrac{z_1(X_{RC}-X_{RC_1})}{1+\cfrac{z_2(X_{RC}-X_{RC_2})}{\vdots\, z_N(X_{RC}-X_{RC_{N-1}})}}},
\end{eqnarray}

where $N$ is a set of input variables (see Fig.3),  and  $z_j$ are the fitting coefficients that  are obtained from  Eq.\ref{SCHLESSINGER} when one substitutes  ${X_{RC}=X_{RC_j}}$ values. As for example $z_1=\big[\frac{V_ad(X_{RC_1})}{V_{ad}(X_{RC_2})}-1\big]/(X_{RC_2}-X_{RC_1})$.

 In the next section, in order to calculate the complex poles of the adiabatic potential, we will carry out an analytical continuation of $V_{ad}(X_{RC})$
into the complex plane, where $X_{RC}\to X_{RC} e^{i\theta}$ (with $\theta \neq 0$). }

The complex resonance poles of the scattering matrix for the adiabatic potential that are embedded above the top of the $V_{c}(X_{RC})$
 potential correspond to resonances arising from energy transfer from the reaction coordinate to the bound bending and symmetric stretching normal modes. This transfer reduces the activation energy required for the reaction from the reactants $D+H_2$ to the products $DH+H$. As will be explained in the next section, the complex resonance pole located at the top of the adiabatic potential barrier determines the rate of the reaction outside the cavity. The cavity couples this resonance with one of the complex resonance poles located below it, thereby enhancing  the reaction rate \color{black}. This explanation is provided in the next section. The results were previously presented in Fig.{\ref{ENHANCEMENT}}.

\section{The complex vibrational resonance poles of the scattering matrix for $D+H2\to DH+H$ and the reaction rate outside of the cavity}

The complex resonance poles (CRP)  of the scattering matrix are the solutions of the time-independent Schr\"odinger equation with outgoing boundary conditions (i.e., zero incoming wave amplitudes)\cite{GAMOW1928,GAMOW2,Siegert1938,NHQM-BOOK}. That is, 
\begin{equation}
\Big( -\frac{\hbar^2}{2m_D}\partial^2_{X_{RC}}+V_{ad}(X_{RC})-E_n^{Com.}\Big)\Psi_n^{Com.}(X_{RC})=0
\label{TISE}
\end{equation}
where the complex eigenvalues is given by,
\begin{equation}
    E_n^{Com.}=E_n-\frac{i}{2}\Gamma_n
\end{equation}
and,
\begin{equation}
    \lim_{X_{RC}\to \pm\infty}\Psi_n^{Com.}(X_{RC})=\sqrt{\frac{M_D}{\hbar^2|k_n|}}e^{\pm k_n^{Com.} X_{RC}}
\end{equation}
where,
\begin{equation}
  k_n^{Com.} X_{RC}=\sqrt{2M_DE_n^{Com.}}/\hbar
\end{equation}
Since $Im[k_n^{CRP}]<0$ the asymptotes of $\Psi_n^{CRP}(X_{RC})$ diverge exponentially. By carrying out the uniform complex scaling transformation  $X_{RC}\to  e^{i\theta}X_{RC}$ the complex scaled eigen functions become square integrable and return to the generalized Hillbert space \cite{NHQM-BOOK}.

By carrying out complex scaling transformation where $X_{RC}\to X_{RC} exp(i\theta)$, and using Eq.\ref{TISE} with the adiabatic potential given in Eq.\ref{SCHLESSINGER}, the spectrum presented in Fig.\ref{COMPLEXpoles} was obtained for $\theta=0.5$. The 
quasi-continuum eigenvalues were rotated by the angle $2\theta=1$.  The resonance complex poles are $\theta$-independent for sufficiently large $\theta$\cite{NHQM-BOOK}. The zero-node complex-scaled resonance function is located at the top of the adiabatic potential shown in Fig.~\ref{1DadiabaticPEC} and is therefore considered the transition state resonance (TSR). 
The reaction outside the cavity occurs when the relative kinetic energy of the reactants is high enough to overcome the potential barrier associated with the transition from reactants to products. 
The imaginary part of the TSR, multiplied by $-2$, gives the decay rate of the chemical reaction $D+H_2\to HD+H$ outside of the cavity. In the next section, we will explain how the cavity couples the TSR with another resonance of lower energy and higher decay rate, thereby enhancing the chemical reaction.

\begin{figure}[h!]
     \centering
     \hspace*{+0.9cm}
      \includegraphics[angle=000,scale=0.5]{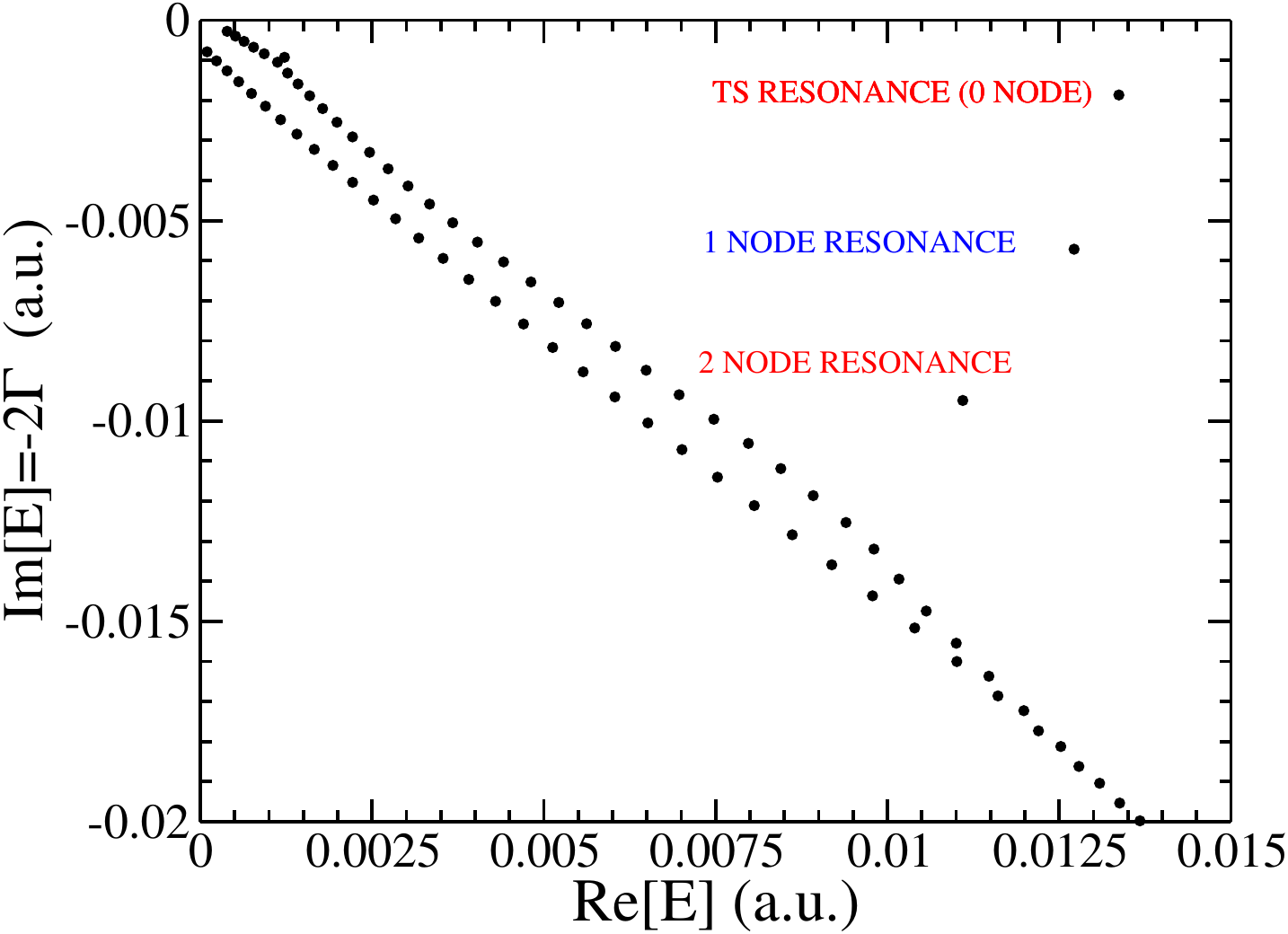}
     \caption{  The resonance complex poles of the scattering matrix for the adiabatic potential shown in Fig.~\ref{1DadiabaticPEC} are displayed. The imaginary parts of these complex resonance poles are associated with the rate of the reaction $D+H_2\to DH+H$ when the total energy of the reactants, including their relative kinetic energy, matches the real parts of the poles. For details on how the poles were calculated, see the text.}
    \label{COMPLEXpoles}
    \end{figure}
    
% %\section{Polariton chemistry: 
% \section{The enhancement of the reaction rate of $D+H_{2}\to DH+H$ inside an initially-dark cavity:
% The case of a single transition-state molecule in the cavity}

\section{Cavity-induced reaction rate enhancement for 
$D+H_{2}\to DH+H$ inside an initially-dark cavity:
The case of a single excited transition-state molecule in the cavity}

Let us use the $|0_{photon}\rangle, |1_{photon}\rangle$ as a basis set for the QED (quantum electrodynamic) waves and the resonance complex Gamow solutions as described in Ref.\citenum{Siegert1938} for the solutions of Eq.\ref{COMPLEXpoles}, $|0_{node}\rangle,|1_{node}\rangle,|2_{node}\rangle$. The polaritons are mixed states of QED and matter waves. The polariton Hamiltonian matrix for $DHH$ in a dark (no light photon) cavity is given by,
\begin{equation}
 H^{(1)}_{polariton}=
\begin{pmatrix}
E_{0_{node}}-0.5i\Gamma_{0_{node}}-\hbar\omega_{cav} & \epsilon_{cav} d_{01} \\
\epsilon_{cav} d_{10} & E_{1_{node}}-0.5i\Gamma_{1_{node}}
\end{pmatrix}
\end{equation}

in the case that $\hbar\omega_{cav}\approx E_{0_{node}}-E_{1_{node}}$. $d_{01}=d_{10}$ are the dipole transitions between the two Gamow-Siegert resonances.
{Note that upon complex scaling the Gamow-Siegert resonance wavefunctions become square integrable.}

The polariton eigenvalues of these Hamiltonian matrices are given by
$E_\pm^{(pol)}-0.5\Gamma_\pm^{(pol)}$ are associated with the polariton wave function,
\begin{equation}    |Pol\rangle_{\pm}=A_\pm|0_{node}\rangle|0_{photon}\rangle +B_\pm|1_{node},2_{node}\rangle|1_{photon}\rangle
\end{equation}
which implies that on resonance condition, where $E_{0_{node}} - \hbar\omega_{cav} = E_{1_{node}}$,
the TS resonance state emit one photon.
Since the initial state is $|0_{node}\rangle$ resonance state (associated with the transition state) the polarition decay rate (i.e., the rate of the reaction inside the cavity) is defined as,
\begin{equation}
    \Gamma_{cav}(\epsilon_{cav})=|A_+|^2\Gamma_++|A_-|^2\Gamma_-^{(pol)}
\end{equation}

In Fig.\ref{ENHANCEMENT} the ratio between the rates inside and outside the cavity is shown as a function of the strength coupling between the molecules and the cavity. 

%----------------------------------------------------------

%\section{Effect of Many Excited Transition-State Molecules in a Dark Cavity on Reaction Rate Enhancement}

\section{Cavity-induced reaction rate enhancement for 
$D+H_{2}\to DH+H$ inside an initially-dark cavity:
The case of many excited transition-state molecules in the cavity}

In the supplementary material (SM)\cite{SM} we present a simple yet insightful theoretical model 
that is designed to investigate how the number of $DHH$ molecules confined within an initiall-dark optical cavity ($N$) influences the rate of a chemical reaction in the gas phase. In this model, it is assumed that the molecules are non-interacting with each other, meaning intermolecular collisions or dipole-dipole interactions are neglected. However, each molecule individually interacts with the quantized electromagnetic field of the cavity. 

Similar to what is presented above, 
for each $DHH$ molecule, the model accounts for two distinct resonance states through which the reaction can take place. 
Initially, only the first one, which is energetically higher, is populated, such that without a cavity it is the only relevant one.
The second resonance state becomes populated only due to the interaction between the molecules and the cavity field,
following the emission of cavity photon(s). 
As a result, for a given pair of such molecular resonance states, 
the reaction rate in a cavity may be enhanced or suppressed 
as compared to the rate without a cavity, 
depending on the number of molecules in the cavity 
and the light-matter coupling inside the cavity.

The SM presents the 
several- and many-molecules results we have calculated for the 
enhancement of the rate of the reaction
D+H$_{2}$$\rightarrow$DH+H 
due to the initially-dark cavity. 
The corresponding single-molecule results 
(i.e., the results with a single pair of reactants in the cavity) are shown above in Fig.~\ref{ENHANCEMENT}. 

%--------------------------------------
%
\begin{figure}[b] %[!] %[b] %[t!] %[b!]
\centering
\includegraphics[angle=000,scale=0.45]
{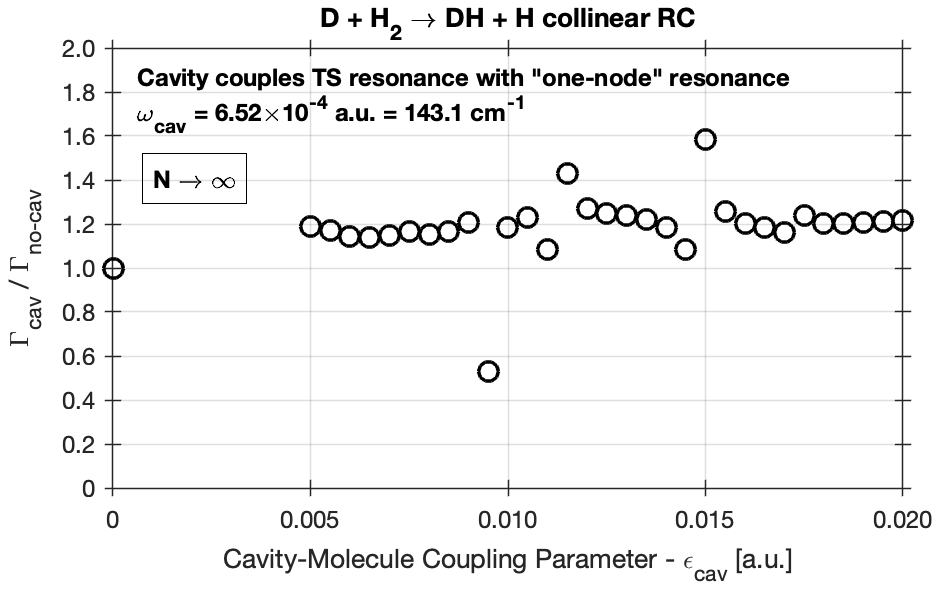}
\caption{ 
Calculated results for the enhancement of the reaction rate due to the initially-dark cavity for the HER reaction D+H$_{2}$$\rightarrow$DH+H  
(see details in the text and the SM~\cite{SM}).
Shown is the cavity-induced rate enhancement,
$\text{R}_{\text{cav,no-cav}} = \Gamma_{\text{cav}} / \Gamma_{\text{no-cav}}$,
as extrapolated in the many-molecule limit of 
infinite number of $DHH$ molecules inside the cavity
(i.e., in the limit of $N$$\rightarrow$$\infty$)
for the values of the coupling parameter between 
the cavity and a single molecule, $\varepsilon_{cav}$, at the range of 0.005 to 0.020 a.u., in steps of 0.0005.
The quantities $\Gamma_{\text{cav}}$ and $\Gamma_{\text{no-cav}}$ denote 
the reaction rates with and without the cavity (i.e, inside and outside the cavity), respectively.
For completeness, the no-cavity point ($\varepsilon_{cav} = 0$), where $\text{R}_{\text{cav,no-cav}} = 1$ by definition, is also included.
See details in the text and the SM~\cite{SM}.
}
\label{paper_Fig_R_vs_epsilon_cav_for_N_inf} 
\end{figure}

%\vspace*{1cm}

%-----------------------------------------------------

For clarity and completeness, % of the paper,
Fig.~4 of the SM, which presents the many-molecule results,
is reproduced here as Fig.\ref{paper_Fig_R_vs_epsilon_cav_for_N_inf}.
It presents the cavity-induced rate enhancement,
$\text{R}_{\text{cav,no-cav}} = \Gamma_{\text{cav}} / \Gamma_{\text{no-cav}}$,
as extrapolated~\cite{SM} in the many-molecule 
limit of 
infinite number of molecules in the cavity 
(i.e., in the limit of $N$$\rightarrow$$\infty$)
for the values of the coupling parameter between 
the cavity and a single molecule, $\varepsilon_{cav}$, at the range of 0.005 to 0.020 a.u., in steps of 0.0005.
The quantities $\Gamma_{\text{cav}}$ and $\Gamma_{\text{no-cav}}$ denote 
the reaction rates with and without the cavity (i.e, inside and outside the cavity), respectively.
For completeness, the no-cavity point ($\varepsilon_{cav} = 0$), where $\text{R}_{\text{cav,no-cav}} = 1$ by definition, is also included.
%
% As seen in Fig.~\ref{paper_Fig_R_vs_epsilon_cav_for_N_inf}, 
% noticeable cavity-induced rate enhancement 
% %
%\text{R}_{\text{cav,no-cav}}(N$$\rightarrow$$\infty)$ 
% is generally obtained 
% in the many-molecule limit 
% across this full range of $\varepsilon_{cav}$.
%
%
As seen in Fig.~\ref{paper_Fig_R_vs_epsilon_cav_for_N_inf}, 
a noticeable cavity-induced rate enhancement 
of about 
$\text{R}_{\text{cav,no-cav}}$(N$\rightarrow$$\infty$)$\sim$1.20 
is generally obtained in the many-molecule limit 
across this full range of $\varepsilon_{cav}$.
However, importantly, at some values of $\varepsilon_{cav}$,
there are prominent deviations from this general value.
%of $\text{R}_{\text{cav,no-cav}}$(N$\rightarrow$$\infty$).
For example, at $\varepsilon_{cav}$=0.015, there is a higher
enhancement of the reaction rate with 
$\text{R}_{\text{cav,no-cav}}$(N$\rightarrow$$\infty$)$\sim$1.60 
while, in contrast, 
at $\varepsilon_{cav}$=0.0095, there is a
an attenuation of the reaction rate with  
$\text{R}_{\text{cav,no-cav}}$(N$\rightarrow$$\infty$)$\sim$0.50. 
This illustrates the extremely high complexity of the system once many molecules 
are coupled to the cavity.
As presented in Fig.~\ref{ENHANCEMENT} above and in 
Fig.~1 of the SM,
the noticeable rate enhancement obtained in the many-molecule case 
is generally similar to the single- and several-molecule cases ($N$=1$-$11).
However, for the single- and several-molecule cases, 
no rate attenuation (i.e., $\text{R}_{\text{cav,no-cav}}$$<$1)
is obtained.
The general characteristic of deviations 
at some values of $\varepsilon_{cav}$
to enhancement values 
higher than the general trend do exist, on the other hand, 
in some of the several-molecule cases 
(for example, for $N$=8 an 10; see SM's Fig.~1).
Overall, our many-molecule results show that, 
in corresponding experiments with the 
HER reaction $D+H_{2}\rightarrow DH+H$ 
conducted along the general scheme presented here,
a noticeable enhancement of the reaction rate due 
to the initially-dark cavity is generally expected.
However, it should also be noted that attenuation of the reaction rate is expected to occur around certain values of the cavity–molecule coupling parameter.
%
% This is similar to the single- and several-molecule cases ($N$=1$-$11),  
% as presented in Fig.~\ref{ENHANCEMENT} above and in the SM.
% %
% So, altogether, our calculations show that,  
% in this range of coupling parameter values, 
% a noticeable enhancement of the reaction rate due to the initially-dark cavity generally exists 
% for the HER reaction D+H$_{2}$$\rightarrow$DH+H 
% in all cases—single-, several-, and many-molecule cases.

%--------------------------------------------------------

%\section{Proposed experiments for studying polariton chemistry in bi-molecular collisions}

\section{Proposed experiments for studying polariton chemistry in gas-phase bimolecular collisions}

The realization of molecular gases at cold and ultracold temperatures enables the study of molecular scattering and reactivity with quantum-state resolution. This opportunity is particularly enticing for the field of cavity control of chemistry, since the cold temperatures enable to remove the detrimental role of thermal excitations and increase the coherence of polaritonic states. In addition, gas-phase reactions do not require solvents and the reaction outcome can be accurately predicted by quantum chemistry calculation. 
Owing to the light mass of the reactants 
and the relative simplicity,
the hydrogen exchange reaction (HER), i.e., $H_2 + H$ and its isotopic variants, such as the $H_2+ D \rightarrow HD + H$ reaction studied throughout this paper, represents an ideal testbed for theory-experiment comparison.

Owing to the limitations of direct laser cooling of hydrogen atoms and molecules, the most suitable experimental approach to study polariton chemistry with HER at low energies is represented by combining cooling and trapping techniques from molecular beam experiments\cite{Meerakker2012}. Recently, merged beam scattering experiments \cite{ED_1,Tang2023} have provided unprecedented resolution in the low-energy scattering of hydrogen isotopologues\cite{Shagam}, thanks to the integration of techniques for reactant detection, such as Velocity-Map Imaging (VMI). In addition, it may be possible to add a cavity in the small interaction region of the merged beams in order to study the effect of strong light-matter coupling on molecular scattering. However, the limited number of collisions per experimental shot combined with the limited time of particles scattering in the cavity volume, makes the investigation of collective effects in polariton chemistry practically impossible. The study of collective effects requires the trapping of molecules at high density inside the optical cavity. 

The best approach for trapping cold molecules inside an optical cavity is represented by Cryogenic Buffer Gas Cells (CBGCs)\cite{Hutzler2012}. CBGCs are an ideal tool for delivering internally cold molecules at high densities and low velocities, even below the capture velocity of conventional conservative traps. In addition, CBGCs can be easily integrated with optical or infrared cavities for cavity-enhanced spectroscopy or molecular control. The Weichman group achieved the first experimental observation of molecular polaritons for a buffer-cooled but gas of methane traveling inside a high-reflectivity infrared cavity\cite{Wright2023}. Recently, infrared cavities integrated with CBGC have emerged as a promising platform for directly trapping different molecular species, including molecular hydrogen H$_2$, at high densities, (e.g., molecular density  $ n > 10^{15}$ cm$^{-3}$)\cite{Singh2023,stankiewicz2025}. The purpose of the cavity is thus twofold: on one hand, we can use it to engineer strong light-molecule coupling when the cavity resonates with specific transitions in the molecular potential or matches HER scattering poles \cite{truong2023mid}. On the other hand, we can simultaneously use the power built up inside the cavity mirrors to high-volume conservative traps, that do not resonate with any transition of the molecular spectrum, nor their higher harmonics. Following the cavity design in Ref.\citenum{Singh2023}, we can generate trapping potential larger than 10 K with mode waist of around $30\ \mu m$, by simply using commercially available infrared lasers with output power of only 1 W. A factor of 4 increase in input power can result in similar trap depth in the confocal regime with improved stability and larger mode volume. The introduction of atomic deuterium at low energies can be accomplished with an additional inlet in the CBGC or by using alternative techniques such as Zeeman deceleration, which can robustly deliver dense atomic beams with thermal velocities well below 1 K, suitable for trapping in the cavity. The combination of infrared cavities, CBGCs, and multistage Zeeman deceleration offers a modular approach for rapidly changing cold reactant species and targeting different or more favorable bimolecular reactions. For example, simply replacing $H_2$ with $HD$ in the CBGC inlet allows the study of the exothermic reaction $HD + HD \rightarrow H_2 + D_2$, which has a lower reaction energy\cite{tomza2015energetics}.
For such a cavity design, we expect a number of trapped molecules and atoms exceeding 100 million counts. 
%
% The collective $\sqrt{N}$-scaling of light-matter interactions with the number of trapped molecules $N$ inside the cavity results in larger cavity-induced effects, including a larger vacuum Rabi splitting and stronger light-matter hybridization. Observing cavity-induced effects in the ground state reaction dynamics is however more complicated. Bi-molecular reactions, as the ones described here, are not a collective phenomena and rather they happen  after independent short-range collisions between two molecules. Since chemistry goes molecule by molecules, it is possible that the effective cavity-induced ground-state correction per molecules becomes vanishing small (i.e., $1/N$-scaling) and collective effects do not affect the overall reaction dynamics, but rather only a narrow energy window of the reaction landscape.
%

 The crucial aspect of our proposed apparatus 
 shown in Fig.~\ref{FigExp} 
 is the possibility to zoom into specific reactions and filter out all the undesired dynamics that can hide cavity-induced effects. The photons emitted in the new type of chemical reaction proposed here, i.e., $D+H_2\to DH+H +\hbar\omega_{photon}$, can act as messengers for the reactions inside the cavity. When photons are emitted, they can be collected by the cavity, leak out from the mirrors after few round trips, and hit a photodiode for detection. The optical signal can be converted into a trigger for then activating the ion optics and imaging with the mass spectrometer. This is the crucial advantage of our experimental scheme: the cavity photon offers a way to remove all the reactions that happen in the background and are not directly related to cavity-induced modifications. We point out that the cavity effectively works as a monochromator that cleans the spectral purity of the optical signal and further filter out the relevant region in reactant detector, allowing us to zoom into the region of the reaction landscape that is energetically accessible. 

\begin{figure}%[h!]
     \centering
    \hspace*{+0.9cm}
      \includegraphics[angle=000,scale=1.0]{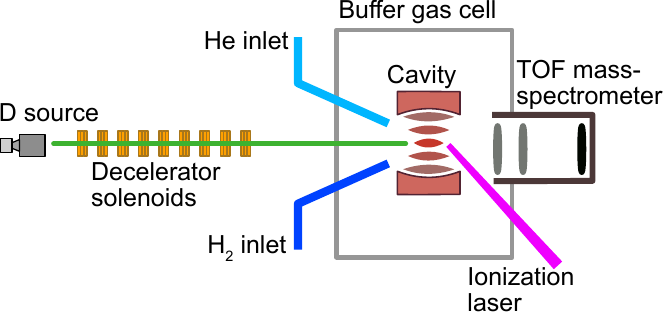}
     \caption{Experimental setup for studying polaritonic chemistry in bi-molecular scattering at cold temperatures. An infrared cavity inside a CBGC can be used to trap cold molecules and probe molecular polaritons. Target molecular species can be introduced by dedicated inlets of deceleration stages. Helium (He) serves as a coolant for molecular internal excitation and improves state preparation of the reactants. A broadband TOF mass-spectrometer can be used to probe the distribution of chemical reactions in combination with state-resolved laser ionization.  }
    \label{FigExp}
\end{figure}

%%--------------------------------------------------------

\section{Concluding remarks}
Studying one of the simplest asymmetric reactions, 
$D + H_2 \to DH + H$, 
we show here that even a weak symmetry breaking in chemical reactions~—~where products differ from reactants due solely to the use of different isotopes~—~is sufficient to enhance the reaction rate within 
an appropriately chosen dark cavity. 
Experimental confirmation of the corresponding cavity-induced reaction pathway and of the many-molecule theoretical results presented here would represent a significant step toward closing the gap between theory and experiment in studies of QED effects on gas-phase reaction rates involving dark cavities.

\vspace*{1cm}

\acknowledgments{Dr. Arie Landau from the Helen Diller Center for Quantum Mechanics at the Technion is acknowledged for most helpful comments. {This research was partially supported by the Israel Science Foundation (ISF) grant No. 1757/24 given to (NM), by the Ministry of Science and the Arts under the KAMEA program given to (VB) and by the European Union (ERC, LIRICO 101115996)} given to (GV).}
\bibliography{References}

\end{document}

%\title{Effect of Many Excited Transition-State Molecules in a Dark Cavity on Reaction Rate Enhancement}

\title{Supplemental Material: Effect of Many Excited Transition-State Molecules in a Dark Cavity on Reaction Rate Enhancement}

\author{Zohar Amitay$^1$\footnote{\tt Correspondence author, amitayz@technion.ac.il}, 
Victor Berenstein$^1$, 
Giacomo Valtolina$^4$ and Nimrod Moiseyev$^{1,2,3}$}

\affiliation{Schulich Faculty of Chemistry, 
$^2$Faculty of Physics and $^3$Solid State Institute, Technion-Israel Institute of Technology, Haifa 32000, Israel,
$^4$Fritz-Haber-Institut der Max-Planck-Gesellschaft, Faradayweg 4-6, 14195 Berlin, Germany
}

\maketitle

The purpose of this appendix is to present a simple yet insightful theoretical model designed to investigate how the number of molecules confined within an initially-dark optical cavity influences the rate of a chemical reaction in the gas phase. In this model, it is assumed that the molecules are non-interacting with each other, meaning intermolecular collisions or dipole-dipole interactions are neglected. However, each molecule individually interacts with the quantized electromagnetic field of the cavity. 
For each molecule, the model accounts for two distinct resonance states through which the reaction can take place. 
Initially, only the first one, which is energetically higher, is populated, such that without a cavity it is the only relevant one.
The second resonance state becomes populated, 
as part of the polaritonic states of the cavity,
only due to the interaction between the molecule and the cavity field, 
along with the corresponding emission of a cavity photon.
As a result, for a given pair of such molecular resonance states, each corresponding to a different individual reaction rate, 
the overall reaction rate in a cavity may be either enhanced or suppressed 
relative to the rate without a cavity, 
depending on the cavity-molecule coupling strength and 
the number of molecules in the cavity. 

%-------------------------------------

The Hamiltonian of $N$ non-interacting 
molecules in a single-mode cavity 
is given within the length gauge and dipole approximation, 
after neglecting the dipole self-energy terms, by 
%
%\begin{align}
\begin{equation}
\hat{H}_{cav,Nmol} = \hbar\omega_{cav}\hat{a}^{\dagger}_{cav}\hat{a}_{cav} + 
                \sum_{i=1}^{N} \left[ 
                \hat{H}_{mol,i} + 
                \varepsilon_{cav} (\hat{a}^{\dagger}_{cav}+\hat{a}_{cav})\hat{d}_{mol,i} 
                \right] , 
                \nonumber
\end{equation}
%
% _\mathrm{cav}  - subscript in regular text
%
where $\hat{H}_{mol,i}$ and $\hat{d}_{mol,i}$ are, respectively, the 
%(cavity-free) 
%complex-scaled 
Hamiltonian and dipole-moment operator of an individual molecule $i$; 
$\omega_{cav}$ and $\varepsilon_{cav}$ are, respectively, the cavity frequency and 
the coupling parameter between the cavity and a single molecule; 
%cavity-molecule coupling parameter;
and $\hat{a}^{\dagger}_{cav}$ and $\hat{a}_{cav}$ are, respectively, the 
creation and annihilation operators of the cavity photonic field.
We assume here, for simplicity, that all the molecules are identical and that they are aligned along the direction of the (single-mode) cavity field.
As described below, in our present model, we apply the rotating-wave approximation and work within a molecular subspace spanned by two molecular states.
Consequently, the above Hamiltonian, $\hat{H}_{\text{cav},N\text{mol}}$, reduces to the so-called Tavis–Cummings Hamiltonian~\cite{TavisCummingsModel}.

% ###
% The eigenvalue of the non-Hermitian complex-scaled Hamiltonian 
% corresponding to each resonance state is complex, 
% where the real part is the energy of the state,
% and the imaginary part is proportional to the rate of the reaction when it takes place via this state.
% ######

As a model, we work 
within the non-Hermitian quantum framework~\cite{NHQM-BOOK},
in the molecular sub-space spanned by the two molecular resonance states 
$\left|\phi\right>_{mol} = \left\{ \left|TS\right>_{mol} , \left|DB\right>_{mol} \right\}$, 
%$ = \left|DB\right>_{mol}$  
%
which have $E^{(mol)}_{TS}$ and $E^{(mol)}_{DB}$ as 
their respective complex eigenvalues of the non-Hermitian complex-scaled Hamiltonian,
where 
$\Re(E^{(mol)}_{TS}) > \Re(E^{(mol)}_{DB}) \,$ 
[$\Re$ denotes the corresponding real part].
For each resonance state, 
the real part of the corresponding complex Hamiltonian eigenvalue 
is the energy of the state,
while the imaginary part is proportional to the rate of the reaction when it takes place via this state.
Each of these resonance states corresponds to a different reaction pathway from the reactants to the products. 
The $\left|TS\right>_{mol}$ resonance state is a conventional reaction transition state, 
where the reaction takes place at an energy close to the height of the reactants-to-products potential barrier.
In contrast, the $\left|DB\right>_{mol}$ resonance state 
%, on the other hand, 
corresponds to a non-conventional reaction pathway, 
in which the reaction takes place  
through a dynamical barrier that is associated with the energy transfer 
(mediated by non-adiabatic couplings) from the reaction coordinate to the bound normal modes %which are 
perpendicular to the reaction coordinate. 
In a collinear reaction A+BC$\to$AB+C, the reaction coordinate is the  asymmetric normal mode, where the normal bound modes 
%that are 
perpendicular to the reaction coordinate are the symmetric and the two bending normal bound 
vibrations~\cite{moiseyev2024conditions}.

The initial state of the system is the collective state 
%$\left|TS_{1} \cdots TS_{N};0\right> = 
$\left( \left|TS\right>_{mol,1}\cdots\left|TS\right>_{mol,N} \right) \left|0\right>_{cav}$, 
where all the $N$ molecules are at $\left|TS\right>_{mol}$ 
and the cavity is dark (i.e., it contains no photons).
So, within the rotating-wave approximation,
the polaritonic subspace relevant to our model 
includes a total of $N$ excitations that are distributed across the molecules and the cavity photonic field.
A basis set spanning this $N$-excitations subspace is composed of $K_{\text{coll.states}}$=2$^{N}$ collective states,
each of which is of the general form 
\begin{equation}
\left|\Psi_{k=1..2^N}\right> = 
\left( \prod_{m_{i}^{(k)}=m_{1}^{(k)}...m_{n_{ph}}^{(k)}} \left| DB \right>^{(k)}_{mol,m_{i}^{(k)}}
\right)
\left( \prod_{m_{i}^{(k)}=m_{n_{ph}+1}^{(k)}...m_{N}^{(k)}} \left| TS \right>^{(k)}_{mol,m_{i}^{(k)}}
\right)
\left|n_{ph}^{(k)}\right>_{cav}
%
%\left( \prod_{i=1}^{N} \left| \phi \right>^{(k)}_{mol,i} \right) %\left|n_{ph}^{(k)}\right>_{cav},
\label{Psi_k}
\end{equation}
where $m_{i}^{(k)} \in \{1,2,...,N\}$ and 
$m_{i}^{(k)} \ne m_{j}^{(k)}$ for $i \ne j$.
%
%To clarify the meaning of Eq.~(\ref{Psi_k}):
The collective state $\left|\Psi_{k}\right>$ corresponds to a situation in which there are 
$n_{ph}^{(k)}$ photons in the cavity,
a sub-group of $(N-n_{ph}^{(k)})$ molecules in the $\left|TS\right>_{mol}$ state,   
and the complementary sub-group of 
$n_{ph}^{(k)}$ molecules in the $\left|DB\right>_{mol}$ state. 
Note that, in general, multiple collective states correspond to the same number of photons in the cavity, with each such state  
corresponding to a different partition of the $N$ molecules between these two sub-groups. 
%a sub-group of $(N-n_{ph}^{(k)})$ molecules in $\left|TS\right>_{mol}$ 
%and the complementary sub-group of 
%$n_{ph}^{(k)}$ molecules in $\left|DB\right>_{mol}$.
%
%and the cavity contains $n_{ph}$ photons, 
%such that the total number of excitations remains $N$.
%
The total number of such collective states is 2$^{N}$ 
because each molecule can be in one of two possible states.
%
%In a collective state containing $N-m$ molecules in $\left|TS\right>_{mol}$ and 
%$m$ molecules in $\left|DB\right>_{mol}$, 
%the number of cavity photons $n_{ph}$ takes the (integer) value of $m$  
%to complement the $N$ excitations.

%For clarity, 
Examples of the corresponding basis set 
are given below, for $N=1,2,3$.
%for $N=1,2,3$ with $K_{\text{coll.states}}=2,4,8$, respectively, 
%
For $N$=1, it includes the following two states: 
\begin{eqnarray}
\left|\Psi_{1} \right> = \left|TS\right>_{mol,1} \left|0\right>_{cav}
\quad , \quad %\nonumber \\
\left|\Psi_{2} \right> = \left|DB\right>_{mol,1} \left|1\right>_{cav} \quad .
\nonumber 
\end{eqnarray}
For $N$=2, it includes the following 4 states: 
\begin{align}
\left|\Psi_{1} \right> = &
\left|TS\right>_{mol,1} \left|TS\right>_{mol,2} \left|0\right>_{cav} 
\quad , %\; \quad 
\nonumber \\
\left|\Psi_{2} \right> = &
\left|DB\right>_{mol,1} \left|TS\right>_{mol,2} \left|1\right>_{cav}
\quad , \; \quad %\nonumber \\
\left|\Psi_{3} \right> = 
\left|TS\right>_{mol,1} \left|DB\right>_{mol,2} \left|1\right>_{cav}
\quad , \nonumber \\
\left|\Psi_{4} \right> = &
\left|DB\right>_{mol,1} \left|DB\right>_{mol,2} \left|2\right>_{cav} \quad .
\nonumber 
\end{align}
Last, for $N$=3, it includes the following 8 states:
\begin{align}
\left|\Psi_{1} \right> = &
\left|TS\right>_{mol,1} \left|TS\right>_{mol,2} \left|TS\right>_{mol,3} \left|0\right>_{cav} 
\quad , %\; \quad 
\nonumber \\
\left|\Psi_{2} \right> = &
\left|DB\right>_{mol,1} \left|TS\right>_{mol,2} \left|TS\right>_{mol,3} \left|1\right>_{cav} 
\quad , \; \quad %\nonumber \\
\left|\Psi_{3} \right> = 
\left|TS\right>_{mol,1} \left|DB\right>_{mol,2} \left|TS\right>_{mol,3} \left|1\right>_{cav} 
\quad , \nonumber \\
\left|\Psi_{4} \right> = &
\left|TS\right>_{mol,1} \left|TS\right>_{mol,2} \left|DB\right>_{mol,3} \left|1\right>_{cav} 
\quad , \nonumber \\
\left|\Psi_{5} \right> = &
\left|DB\right>_{mol,1} \left|DB\right>_{mol,2} \left|TS\right>_{mol,3} \left|2\right>_{cav} 
\quad , \; \quad %\nonumber \\
\left|\Psi_{6} \right> = 
\left|DB\right>_{mol,1} \left|TS\right>_{mol,2} \left|DB\right>_{mol,3} \left|2\right>_{cav} 
\quad , \nonumber \\
\left|\Psi_{7} \right> = &
\left|TS\right>_{mol,1} \left|DB\right>_{mol,2} \left|DB\right>_{mol,3} \left|2\right>_{cav} 
\quad , \nonumber \\
\left|\Psi_{8} \right> = &
\left|DB\right>_{mol,1} \left|DB\right>_{mol,2} \left|DB\right>_{mol,3} \left|3\right>_{cav} 
\quad . \nonumber 
\end{align}

%-------------------------------

%Assuming, for simplicity, that $\hbar\omega_{cav} = E^{(mol)}_{TS} - E^{(mol)}_{DB}$, 
%
The $K_{\text{coll.states}} \times K_{\text{coll.states}}$ 
(i.e., $2^{N} \times 2^{N}$)
matrix representation of the non-Hermitian  
Hamiltonian is then given as follows.
The diagonal matrix elements are equal to 
%
%\begin{equation} 
%\left[ \hat{H}_{cav,Nmol}\right]_{k,k} = N E_{TS}^{(mol)} \; .
%\nonumber
%\end{equation}
%
\begin{align} 
\left[ \hat{H}_{cav,Nmol}\right]_{k,k} = & \;
(N-n_{ph}^{(k)}) E_{TS}^{(mol)} + 
n_{ph}^{(k)} E_{DB}^{(mol)} + 
n_{ph}^{(k)} \hbar \omega_{cav} \; =
\nonumber \\
%
%= & \; N E_{TS}^{(mol)} - 
%n_{ph}^{(k)} (E_{TS}^{(mol)} - E_{DB}^{(mol)} - \hbar \omega_{cav}) 
%\nonumber \\
%
= & \; N E_{TS}^{(mol)} - 
n_{ph}^{(k)} \left[ 
E_{TS}^{(mol)} - \left( E_{DB}^{(mol)} + \hbar \omega_{cav}\right) 
\right]
\nonumber 
\end{align}
The non-diagonal matrix element between a pair of collective states $k_{1}$ and $k_{2}$
%$\left|\Psi_{k_{1}}\right>$ and $\left|\Psi_{k_{2}}\right>$ 
is non-zero only for 
$n_{ph}^{(k_{2})} = n_{ph}^{(k_{1})} \pm 1$ 
and when 
$\left|\Psi_{k_{1}}\right>$ and $\left|\Psi_{k_{2}}\right>$ 
are different one from the other in the state of only a single molecule.
In such a case, it is equal to 
\begin{align}
\left[ \hat{H}_{cav,Nmol}\right]_{k_{1},k_{2}} = 
\varepsilon_{cav} \cdot d_{k1,k2} \cdot \sqrt{\text{max}\left(n_{ph}^{(k_{1})},n_{ph}^{(k_{2})}\right)}
\cdot \delta_{n_{ph}^{(k_{1})},n_{ph}^{(k_{2})}\pm1}
\nonumber
\end{align}
where, within the non-Hermitian formalism,  
$d_{k1,k2}$ = 
$d_{DB,TS}$ = 
$^{\ast}$$\left<DB\right|\hat{d}_{mol}\left|TS\right>$
or 
$d_{k1,k2}$ =
$d_{TS,DB}$ = 
$^{\ast}$$\left<TS\right|\hat{d}_{mol}\left|DB\right>$ 
according to the specific pair of states $k_{1}$ and $k_{2}$. 
%
%where 
We have assumed here that 
$_{mol,i}$$^{*}$$\left<DB\right|\hat{d}_{mol,i}\left|TS\right>_{mol,i}$ = $^{*}$$\left<DB\right|\hat{d}_{mol}\left|TS\right>$ 
and
$_{mol,i}$$^{*}$$\left<TS\right|\hat{d}_{mol,i}\left|DB\right>_{mol,i}$ = $^{*}$$\left<TS\right|\hat{d}_{mol}\left|DB\right>$ 
for any molecule $i$.
In these matrix elements, we use the bi-orthogonal product (so called c-product) rather than the scalar product.

%----------------------------------

Examples of the corresponding Hamiltonian are given below,
for $N=1,2$.
For $N=1$, it is given by 
\begin{equation}
\hat{H}_{cav,1\,mol} = 
\begin{pmatrix}
E_{TS}^{(mol)} & 
\varepsilon_{cav} \cdot d_{DB,TS} \cdot \sqrt{1} 
\\
\varepsilon_{cav} \cdot d_{TS,DB} \cdot \sqrt{1}  & 
E_{DB}^{(mol)}+\hbar\omega_{cav}
\end{pmatrix}
. \nonumber
\end{equation}
%
%\begin{equation}
%\hat{H}_{cav,1mol} = 
%\begin{pmatrix}
%E_{TS}^{(mol)} & 
%\varepsilon_{cav} \cdot %\left<DB\right|\hat{d}_{mol}\left|TS\right> \ \cdot \sqrt{1} 
%\\
%\varepsilon_{cav} \cdot %\left<TS\right|\hat{d}_{mol}\left|DB\right> \cdot \sqrt{1}  & 
%E_{DB}^{(mol)}+\hbar\omega_{cav}
%\end{pmatrix}
%\nonumber
%\end{equation}
%
For $N=2$, it is given by 
\begin{align}
& \hat{H}_{cav,2\,mol} = \nonumber \\
& = \begin{pmatrix}
2 E_{TS}^{(mol)} \;\; & 
\varepsilon_{cav} \cdot d_{DB,TS} \cdot \sqrt{1} & 
\varepsilon_{cav} \cdot d_{DB,TS} \cdot \sqrt{1} & 
0 
\\
\varepsilon_{cav} \cdot d_{TS,DB} \cdot \sqrt{1} & 
E_{TS}^{(mol)} + E_{DB}^{(mol)} + \hbar\omega_{cav} & 
0 & 
\varepsilon_{cav} \cdot d_{DB,TS} \cdot \sqrt{2} 
\\
\varepsilon_{cav} \cdot d_{TS,DB} \cdot \sqrt{1} & 
0 &
E_{TS}^{(mol)} + E_{DB}^{(mol)} + \hbar\omega_{cav} & 
\varepsilon_{cav} \cdot d_{DB,TS} \cdot \sqrt{2} & 
\\
0 & 
\varepsilon_{cav} \cdot d_{TS,DB} \cdot \sqrt{2} & 
\varepsilon_{cav} \cdot d_{TS,DB} \cdot \sqrt{2} & 
2 E_{DB}^{(mol)} + 2 \hbar\omega_{cav}
%
%E_{TS}^{(mol)} & \varepsilon_{cav} \cdot \left<DB\right|\hat{d}_{mol}\left|TS\right> \ \cdot \sqrt{1} \\
%\varepsilon_{cav} \cdot \left<TS\right|\hat{d}_{mol}\left|DB\right> \cdot \sqrt{1}  & E_{DB}^{(mol)}+\hbar \omega_{cav}} 
\end{pmatrix}
. \nonumber
\end{align}
%
%\begin{align}
%& \hat{H}_{cav,2mol} = \nonumber \\
%& = \begin{pmatrix}
%2 E_{TS}^{(mol)} \;\; & 
%\varepsilon_{cav} \left<DB\right|\hat{d}_{mol}\left|TS\right> \sqrt{1} & 
%\varepsilon_{cav} \left<DB\right|\hat{d}_{mol}\left|TS\right> \sqrt{1} & 
%0 
%\\
%\varepsilon_{cav} \left<TS\right|\hat{d}_{mol}\left|DB\right> \sqrt{1} & 
%E_{TS}^{(mol)} + E_{DB}^{(mol)} + \hbar\omega_{cav} & 
%0 & 
%\varepsilon_{cav} \left<DB\right|\hat{d}_{mol}\left|TS\right> \sqrt{2} 
%\\
%\varepsilon_{cav} \left<TS\right|\hat{d}_{mol}\left|DB\right> \sqrt{1} & 
%0 &
%E_{TS}^{(mol)} + E_{DB}^{(mol)} + \hbar\omega_{cav} & 
%\varepsilon_{cav} \left<DB\right|\hat{d}_{mol}\left|TS\right> \sqrt{2} 
%\\
%0 & 
%\varepsilon_{cav} \left<TS\right|\hat{d}_{mol}\left|DB\right> \sqrt{2} & 
%\varepsilon_{cav} \left<TS\right|\hat{d}_{mol}\left|DB\right> \sqrt{2} & 
%2 E_{DB}^{(mol)} + 2 \hbar\omega_{cav}
%%
%%E_{TS}^{(mol)} & \varepsilon_{cav} \cdot \left<DB\right|\hat{d}_{mol}\left|TS\right> \ \cdot \sqrt{1} \\
%%\varepsilon_{cav} \cdot \left<TS\right|\hat{d}_{mol}\left|DB\right> \cdot \sqrt{1}  & E_{DB}^{(mol)}+\hbar \omega_{cav}} 
%\end{pmatrix}
%\nonumber
%\end{align}
%

The 2$^{N}$ eigenstates of $\hat{H}_{cav,Nmol}$ are the so-called polaritonic eigenstates of the system, 
each of which has a corresponding eigenfunction 
$\left|\Phi_{P=1\cdots2^{N}}\right>$ that is a superposition of 
the different collective states $\left|\Psi_{k=1\cdots2^{N}}\right>$, 
with corresponding complex coefficients $C_{P,k}$.
The population of a given polaritonic eigenstate $P$
is equal to $\text{P}_{P}$=$\left|C_{P,k=1}\right|^{2}$,
where $C_{P,k=1}$ is the
%the square of the absolute value of the 
$P$-state's superposition coefficient 
corresponding to the initial state of the system, 
%i.e., the collective state 
$\left|\Psi_{k=1}\right>$=$\left( \left|TS\right>_{mol,1}\cdots\left|TS\right>_{mol,N} \right) \left|0\right>_{cav}$.
The $\hat{H}_{cav,Nmol}$'s eigenvalue $E_{P}$ 
corresponding to a given polaritonic eigenstate is 
generally complex, 
with the rate of the reaction taking place via this polaritonic eigenstate given by $\Gamma_{P} = -2 \, \Im(E_{P})$, 
where $\Im(E_{P})$ denotes the imaginary part of $E_{P}$.
Hence, the total reaction rate in the cavity is given by
$\Gamma_{\text{cav}} = \sum_{P=1}^{2^{N}} \text{P}_{P} \cdot \Gamma_{P}$.

%==================###======================

Following the above formulation, 
we have numerically calculated the enhancement of the reaction rate due to 
the initially-dark cavity for the HER reaction D+H$_{2}$$\rightarrow$DH+H 
with the pair of resonance states 
$\left|TS\right>_{mol}$ and $\left|DB\right>_{mol}$,
for which the results are given in Fig.~1 of the paper.
The corresponding values of the different quantities involved in the calculations are
$E^{(mol)}_{TS}$=$1.34\times10^{-2} - 1.87\times10^{-3} i$ a.u., 
$E^{(mol)}_{DB}$=$1.27\times10^{-2} - 5.71\times10^{-3} i$ a.u., 
$d_{TS,DB}$=$0.12 + 0.14 i$ a.u.,  
and
$\omega_{cav}$=$\Re(E^{(mol)}_{TS}) - \Re(E^{(mol)}_{DB})$=$6.52\times10^{-4}$ a.u.=143.1 cm$^{-1}$ 
(with the "a.u." stands for "atomic units").
%All these values, as well as the ones given below, 
%are given in atomic units.
%
Denoting the reaction rates with and without the initially-dark cavity
(i.e, inside and outside the cavity)
by $\Gamma_{\text{cav}}$ and $\Gamma_{\text{no-cav}}$, respectively, 
the cavity-induced rate enhancement is given by 
$\text{R}_{\text{cav,no-cav}} = \Gamma_{\text{cav}} / \Gamma_{\text{no-cav}}$.
%
% As the initial state of the system is
% $\left( \left|TS\right>_{mol,1}\cdots\left|TS\right>_{mol,N} \right) \left|0\right>_{cav}$,
% an upper limit for $\text{R}_{\text{cav,no-cav}}$,
% independent of $N$,
% is given by the ratio $\Gamma_{DB} / \Gamma_{TS}$.
% It is equal here to 3.05.

\begin{figure} %[b!]
\includegraphics[angle=000,scale=0.40,left]
{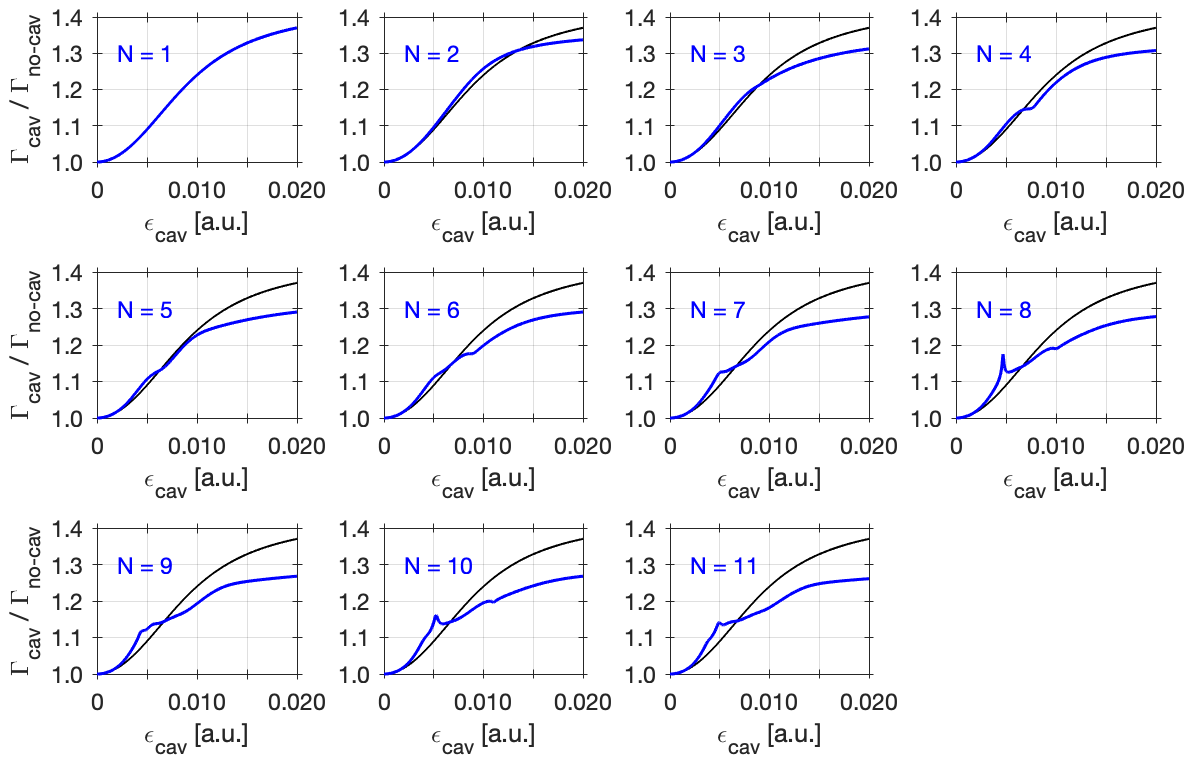}
\caption{ 
Calculated results for the enhancement of the HER reaction rate due to the 
initially-dark cavity for the reaction D+H$_{2}$$\rightarrow$DH+H 
(see details in the text).
Shown (in blue) %circles) 
is the cavity-induced rate enhancement,
$\text{R}_{\text{cav,no-cav}} = \Gamma_{\text{cav}} / \Gamma_{\text{no-cav}}$,
as a function of the coupling parameter between the cavity and a single molecule, $\varepsilon_{cav}$, 
for different numbers of molecules in the cavity, $N$.
The results for $N$=1 are shown (in black), for comparison, 
in all the panels of the figure. %(solid black lines).
The quantities $\Gamma_{\text{cav}}$ and $\Gamma_{\text{no-cav}}$ denote 
the reaction rates with and without the cavity (i.e, inside and outside the cavity), respectively. 
See details in the text.
}
\label{Fig_R_vs_espilon_cav_for_different_N}
\end{figure}

\begin{figure} %[b!]
\includegraphics[angle=000,scale=0.43,left]
{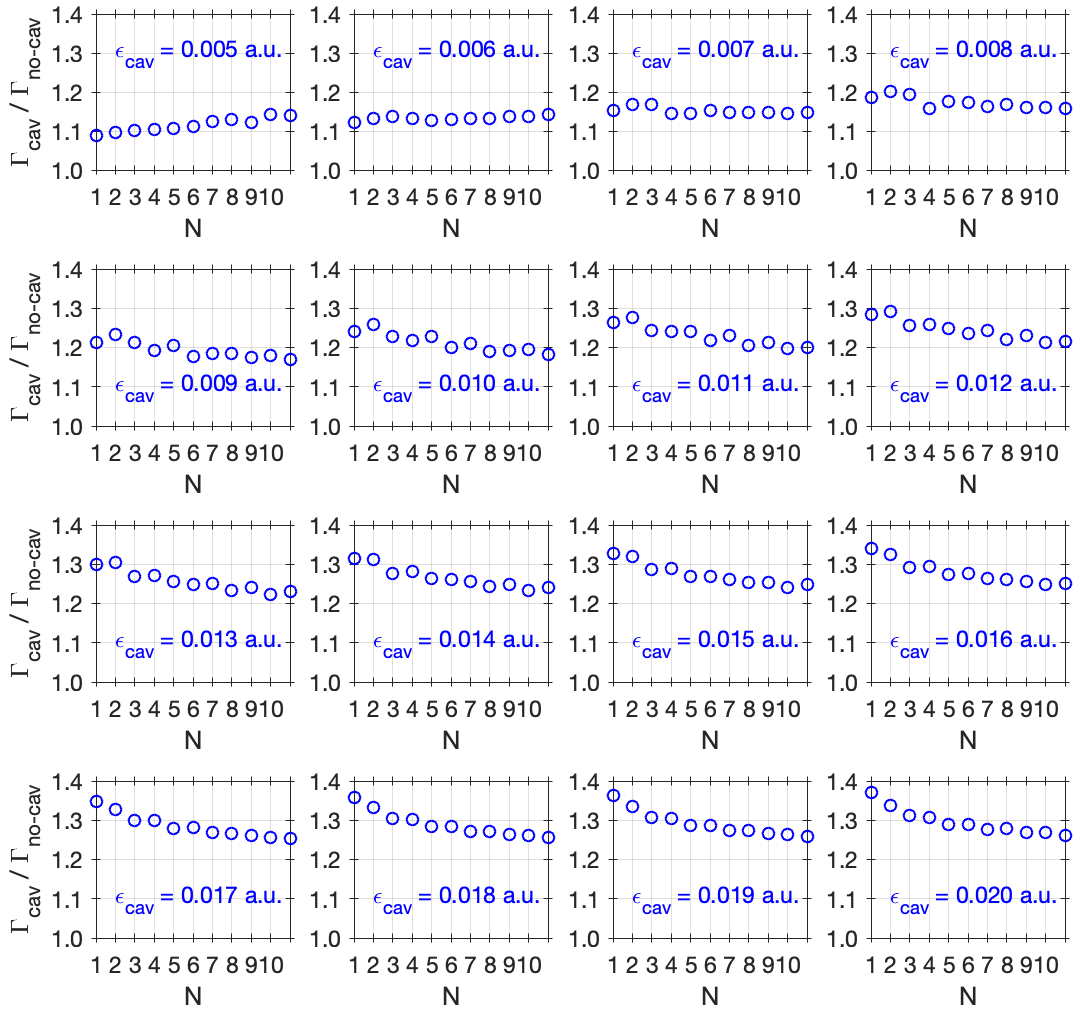}
\caption{
Calculated results for the enhancement of the HER reaction rate due to the 
initially-dark cavity for the reaction D+H$_{2}$$\rightarrow$DH+H 
(see details in the text).
Shown is the cavity-induced rate enhancement,
$\text{R}_{\text{cav,no-cav}} = \Gamma_{\text{cav}} / \Gamma_{\text{no-cav}}$,
as a function of the 
number of molecules in the cavity, $N$, 
for several %different 
representative values of the 
coupling parameter between the cavity and a single molecule, $\varepsilon_{cav}$. 
The quantities $\Gamma_{\text{cav}}$ and $\Gamma_{\text{no-cav}}$ denote 
the reaction rates with and without the cavity (i.e, inside and outside the cavity), respectively.
See details in the text.
}
\label{Fig_R_vs_N_for_different_epsilon_cav}
\end{figure}

Figures~\ref{Fig_R_vs_espilon_cav_for_different_N} 
and \ref{Fig_R_vs_N_for_different_epsilon_cav} present results of $\text{R}_{\text{cav,no-cav}}$
calculated for different numbers of molecules in the cavity, $N$, 
with different values of the coupling parameter between the cavity and a single molecule, $\varepsilon_{cav}$.
The values of $N$ range from 1 to 11, 
as set by our numerical computational capabilities,
%(as our computational capabilities allow), 
and the values of $\varepsilon_{cav}$ range from 0 to 0.020 a.u. %, in steps of 0.00005.
The calculated values of $\text{R}_{\text{cav,no-cav}}$ as a function of $\varepsilon_{cav}$ for the different given values of $N$ 
are shown (in blue) in Fig.~\ref{Fig_R_vs_espilon_cav_for_different_N},
while the calculated values of $\text{R}_{\text{cav,no-cav}}$ as a function of $N$ for different example values of $\varepsilon_{cav}$ 
are shown (in blue) in Fig.~\ref{Fig_R_vs_N_for_different_epsilon_cav}.
In Fig.~\ref{Fig_R_vs_espilon_cav_for_different_N}, 
the results for $N$=1 are also shown (in black), for comparison, 
in all the panels of the figure. 
Note that these $N$=1 results are identical to those presented in Fig.~1 of the paper.
As seen from the figures,  
%Fig.~\ref{Fig_R_vs_N_for_different_epsilon_cav},
for $0.007\le \epsilon_{cav}\le 0.020$ a.u.,
$\text{R}_{\text{cav,no-cav}}$ 
generally decreases as $N$ increases,
whereas for $0.005\le \epsilon_{cav}\le 0.006$ a.u., the 
trend is reversed.
For $\epsilon_{cav}$$<$0.005, 
negligible changes in $\text{R}_{\text{cav,no-cav}}$
are obtained over the present range of $N$=1$-$11.

%From the results of our calculations presented in Fig.\ref{Fig_R_vs_N_for_different_epsilon_cav} one can see that in the interval of $0.005\le \epsilon_{cav}\le 0.02$ by increasing the number of molecules inside the cavity (and the number of emitted photons) the ratio $\Gamma_{cav}/\Gamma_{no-cav}$ is reduced.

%------------------

\begin{figure} %[b!]
\centering
%\hspace*{+0.9cm}
\includegraphics[angle=000,scale=0.55]{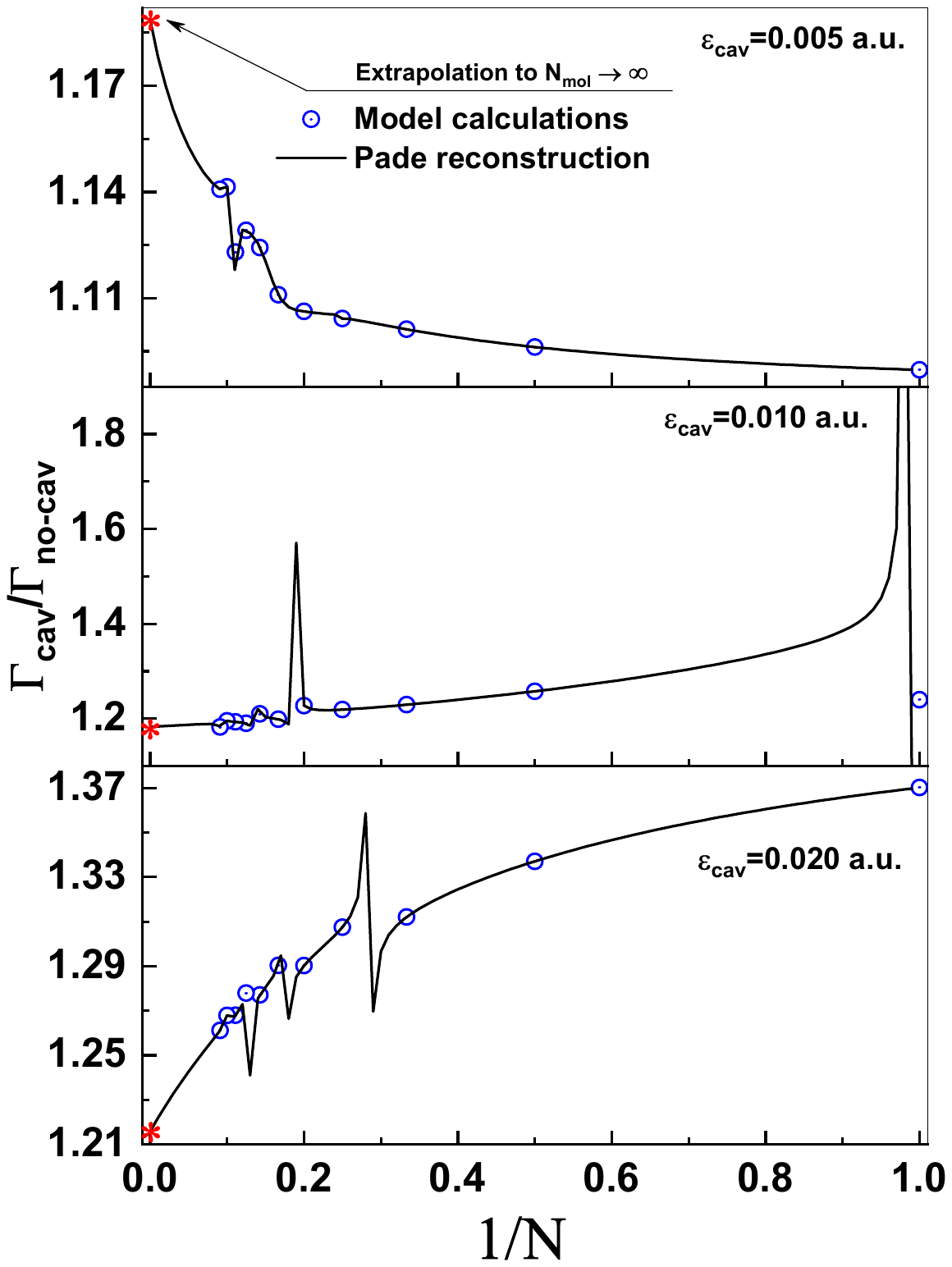}
\caption{
Calculated results for the enhancement of the HER reaction rate due to the 
initially-dark cavity for the reaction D+H$_{2}$$\rightarrow$DH+H 
(see details in the text).
Shown is the cavity-induced rate enhancement,
$\text{R}_{\text{cav,no-cav}} = \Gamma_{\text{cav}} / \Gamma_{\text{no-cav}}$,
as a function of $1/N$, 
where $N$ is the number of molecules in the cavity, 
for the example values of 
0.005, 0.010 and 0.020 a.u.~of %~values 
the coupling parameter between the cavity and a single molecule,
$\varepsilon_{cav}$. 
The quantities $\Gamma_{\text{cav}}$ and $\Gamma_{\text{no-cav}}$ denote 
the reaction rates with and without the cavity (i.e, inside and outside the cavity), respectively.
The numerically calculated values of $\text{R}_{\text{cav,no-cav}}$ 
for $N$=1$-$11 are shown in circles,
while the Schlesinger–Padé reconstruction results 
of $\text{R}_{\text{cav,no-cav}}$ are shown in solid lines.
The former are those shown also in Figs.~\ref{Fig_R_vs_espilon_cav_for_different_N} and 
\ref{Fig_R_vs_N_for_different_epsilon_cav}.
As seen, the extrapolated value of $\text{R}_{\text{cav,no-cav}}$ in the limit of 
$N$$\rightarrow$$\infty$ (i.e., $1/N$$\rightarrow$0), indicated by red stars in the three panels, is approximately 1.20 in all the cases presented.
See details in the text.
}
\label{Fig_R_vs_1_over_N_for_different_epsilon_cav}
\end{figure}

In order to obtain results for the cavity-induced rate enhancement 
$\text{R}_{\text{cav,no-cav}}$ in the many-molecule 
limit of $N$$\rightarrow$$\infty$,
we have applied the Schlesinger–Padé reconstruction (as described in the paper) for each set of  
the (numerically) calculated $\text{R}_{\text{cav,no-cav}}$ 
as a function of $N$=1$-$11 for different given values of $\varepsilon_{cav}$.
Technically, the $N$$\rightarrow$$\infty$ limit has been approached by 
applying the Schlesinger–Padé reconstruction to the calculated  
$\text{R}_{\text{cav,no-cav}}$ results as a function of $1/N$ 
(rather than as a function of $N$),
and extrapolating the results to the limit of 
$1/N$$\rightarrow$0.
The Schlesinger–Padé approach allows us to bypass singularities that signal a transition between different analytical behaviors of the nonlinear function $\text{R}_{\text{cav,no-cav}} = \Gamma_{\text{cav}} / \Gamma_{\text{no-cav}}$.
Figure~\ref{Fig_R_vs_1_over_N_for_different_epsilon_cav} shows examples 
of corresponding results of $\text{R}_{\text{cav,no-cav}}$ 
as a function of $1/N$, for 
$\varepsilon_{cav}$=0.005, 0.010 and 0.020 a.u..
The numerically calculated values of $\text{R}_{\text{cav,no-cav}}$ 
for $N$=1$-$11 are shown in circles,
while the Schlesinger–Padé reconstruction results 
of $\text{R}_{\text{cav,no-cav}}$ are shown in solid lines.
The former are those shown also in Figs.~\ref{Fig_R_vs_espilon_cav_for_different_N} and 
\ref{Fig_R_vs_N_for_different_epsilon_cav}.
As seen in the figure, the extrapolated value of 
$\text{R}_{\text{cav,no-cav}}$ in the limit of 
$N$$\rightarrow$$\infty$ (i.e., $1/N$$\rightarrow$0), 
indicated by red stars in the three figure panels, 
is approximately 1.20 in all the cases presented.
Notably, this applies even though 
the results for $\varepsilon_{cav}$=0.005 a.u.~show $\text{R}_{\text{cav,no-cav}}$ that increases as $N$ increases (i.e., as $1/N$ decreases), 
whereas 
the results for $\varepsilon_{cav}$=0.010 and 0.020 a.u.~show $\text{R}_{\text{cav,no-cav}}$ that decreases as $N$ increases (i.e, as $1/N$ decreases).
%
% \sout{
% The singularities obtained in the Schlesinger–Padé results indicate 
% the high sensitivity of the cavity-induced rate enhancement,
% $\text{R}_{\text{cav,no-cav}}$,
% to the number of molecules $N$ inside a given cavity
% at some regions of $N$ values.
% }

%----------

\begin{figure} %[b!]
\centering
\hspace*{0.55cm}
\includegraphics[angle=000,scale=0.45,left]
{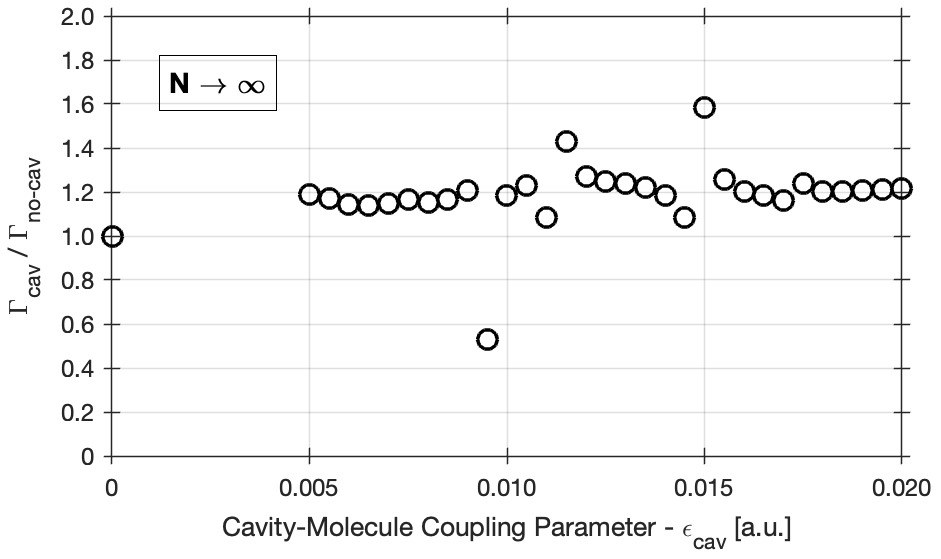}
\caption{
Calculated results for the enhancement of the HER reaction rate due to the 
initially-dark cavity for the reaction D+H$_{2}$$\rightarrow$DH+H 
(see details in the text).
Shown is the cavity-induced rate enhancement,
$\text{R}_{\text{cav,no-cav}} = \Gamma_{\text{cav}} / \Gamma_{\text{no-cav}}$,
as extrapolated in the many-molecule limit of $N$$\rightarrow$$\infty$
by the Schlesinger–Padé reconstruction (see text for details),
for all the values of the coupling parameter between 
the cavity and a single molecule, $\varepsilon_{cav}$, at the 
range of 0.005 to 0.020 a.u., in steps of 0.001.
The quantities $\Gamma_{\text{cav}}$ and $\Gamma_{\text{no-cav}}$ denote 
the reaction rates with and without the cavity (i.e, inside and outside the cavity), respectively.
For completeness, the no-cavity point ($\varepsilon_{cav} = 0$), where $\text{R}_{\text{cav,no-cav}} = 1$ by definition, is also included.
For $\varepsilon_{cav}$=0.005, 0.010 and 0.020 a.u., 
the values presented here are the ones indicated in  Fig.~\ref{Fig_R_vs_1_over_N_for_different_epsilon_cav} by the red stars.
See details in the text.
} 
\label{Fig_R_vs_epsilon_cav_for_N_inf} 
\end{figure}

Figure~\ref{Fig_R_vs_epsilon_cav_for_N_inf} shows
the full set of extrapolated $\text{R}_{\text{cav,no-cav}}$ values 
in the many-molecule limit of $N$$\rightarrow$$\infty$
(obtained as described above using the Schlesinger–Padé reconstruction) 
for all the $\varepsilon_{cav}$ values at the 
range of 0.005 to 0.020 a.u., in steps of 0.0005.
For completeness, we also include the point corresponding to $\varepsilon_{cav} = 0$, i.e., the no-cavity case, where $\text{R}_{\text{cav,no-cav}} = 1$ by definition.
For $\varepsilon_{cav}$=0.005, 0.010 and 0.020 a.u., 
the presented values are the ones indicated in  Fig.~\ref{Fig_R_vs_1_over_N_for_different_epsilon_cav} by the red stars.
As seen, 
a noticeable cavity-induced rate enhancement 
of about 
$\text{R}_{\text{cav,no-cav}}$(N$\rightarrow$$\infty$)$\sim$1.20 
is generally obtained in the many-molecule limit 
across this full range of $\varepsilon_{cav}$.
However, importantly, at some values of $\varepsilon_{cav}$,
there are prominent deviations from this general value.
%of $\text{R}_{\text{cav,no-cav}}$(N$\rightarrow$$\infty$).
For example, at $\varepsilon_{cav}$=0.015, there is a higher
enhancement of the reaction rate with 
$\text{R}_{\text{cav,no-cav}}$(N$\rightarrow$$\infty$)$\sim$1.60 
while, in contrast, 
at $\varepsilon_{cav}$=0.0095, there is a
an attenuation of the reaction rate with  
$\text{R}_{\text{cav,no-cav}}$(N$\rightarrow$$\infty$)$\sim$0.50. 
This illustrates the extremely high complexity of the system once many molecules 
are coupled to the cavity.
As presented in Fig.~\ref{Fig_R_vs_espilon_cav_for_different_N} above (and in Fig.~1 of the paper),
the noticeable rate enhancement obtained in the many-molecule case 
is generally similar to the single- and several-molecule cases ($N$=1$-$11).
However, for the single- and several-molecule cases, 
no rate attenuation (i.e., $\text{R}_{\text{cav,no-cav}}$$<$1)
is obtained.
The general characteristic of deviations 
at some values of $\varepsilon_{cav}$
to enhancement values 
higher than the general trend do exist, on the other hand, 
in some of the several-molecule cases 
(for example, for $N$=8 and 10; see SM's Fig.~1).
Overall, our many-molecule results show that, 
in corresponding experiments with the 
HER reaction $D+H_{2}\rightarrow DH+H$ 
conducted along the general scheme presented in the paper,
a noticeable enhancement of the reaction rate due 
to the initially-dark cavity is generally expected.
However, it should also be noted that attenuation of the reaction rate is expected to occur around certain values of the cavity–molecule coupling parameter.

% As seen, noticeable cavity-induced rate enhancement $\text{R}_{\text{cav,no-cav}}(N$$\rightarrow$$\infty)$ 
% is generally 
% obtained across this full range of $\varepsilon_{cav}$.
% %
% %with substantial variation in its value over certain ranges of $\varepsilon_{cav}$. 
% %
% %
% %with only one exception in the presented points, 
% %at $\varepsilon_{cav}$=0.011 a.u..
% %
% This is similar to the single- and few-molecule cases ($N$=1$-$11),  
% as presented in Figure~\ref{Fig_R_vs_espilon_cav_for_different_N}. 
% %
% So, altogether, 
% in this range of coupling parameter values, 
% the calculations show that 
% a noticeable enhancement of the reaction rate due to the initially-dark cavity generally exists 
% for the HER reaction D+H$_{2}$$\rightarrow$DH+H 
% in all cases—single-, several-, and many-molecule cases.
% %
% %with substantial changes in the enhancement over certain ranges of the cavity–molecule coupling.

%---------

% Experimental confirmation of the many-molecule theoretical results obtained here would be a significant step toward closing the gap 
% between theory and experiment in studies of QED effects on reaction rates in the gas phase involving initially-dark cavities. 

%==========================

\bibliography{References}

\newpage 

%----------

% {app-fig1b-rslts_all_polaritons--eps-cav-step-0.0005-part1.png}
%
% {app-fig1b-rslts_all_polaritons--eps-cav-step-0.0005-part2.png}

%----------

%-------------

%---------------

